\documentclass[compsoc, conference, a4paper, 10pt, times]{IEEEtran}

\usepackage[table]{xcolor}
\usepackage[disable]{todonotes}
\newcommand{\aslantodo}[1]{}
\usepackage[utf8]{inputenc}
\usepackage{microtype}
\microtypecontext{spacing=nonfrench}
\usepackage{amsmath}
\usepackage[hidelinks,breaklinks]{hyperref}

\usepackage{cleveref}
\usepackage{tikz}
\usepackage{listings}

\usepackage{mfirstuc}
\usepackage{tabularx}
\usepackage{booktabs}
\usepackage{soul}

\usepackage{graphicx}
\usepackage{xspace}
\usepackage{enumitem}
\usepackage{amssymb}
\usepackage{microtype}
\usepackage{mathpartir}
\usepackage{amsthm}
\usepackage{framed}
\usepackage{caption}
\usepackage{subcaption}
\usepackage{placeins} 

\usepackage[parfill]{parskip}
\usepackage{stackengine} 

\definecolor{lightgray}{gray}{0.9}
\let\oldtabularx\tabularx
\renewcommand*{\tabularx}{\rowcolors{0}{}{lightgray}\oldtabularx}
\usepackage{array}

\makeatletter
\@ifundefined{c@rownum}{%
  \let\c@rownum\rownum
}{}
\@ifundefined{therownum}{%
  \def\therownum{\@arabic\rownum}%
}{}
\makeatother

\usepackage{pifont}
\newcommand{\cmark}{Yes}%
\newcommand{\xmark}{No}%
\newcommand{\noguarantee}{--}%

\definecolor{orangeish}{HTML}{f1a340} 
\definecolor{grayish}{HTML}{f7f7f7} 
\definecolor{purpleish}{HTML}{998ec3} 

\definecolor{yellowish}{HTML}{DDAA33}
\definecolor{redish}{HTML}{BB5566}
\definecolor{blueish}{HTML}{004488}

\newcommand{\basicCodeStyle}{\ttfamily\small}
\newcommand{\keywordCodeStyle}{\bfseries}
\newcommand{\builtInCodeStyle}{\itshape\color{blueish}}
\newcommand{\typesCodeStyle}{\ttfamily\small\color{blueish}}

\lstdefinelanguage{TypeScript}{
  classoffset=0,
  morekeywords={typeof, new, true, false, catch, function, return, null, catch, switch, var, if, in, while, do, else, case, break, let, for, const, class, implements, enum, get, this, number},
  classoffset=1, 
  morekeywords={NetworkNode, string,EncryptedString,PaddedMsg,MessageType,DummyString},
  keywordstyle=\typesCodeStyle,
  classoffset=2, 
  morekeywords={getPadBytes,size},
  keywordstyle=\builtInCodeStyle,
  classoffset=0,
  morecomment=[s]{/*}{*/},
  morecomment=[l]//,
  morestring=[b]",
  morestring=[b]'
}

\lstdefinelanguage{Protobuf}{
  classoffset=0,
  morekeywords={message, required, repeated, optional},
  classoffset=1, 
  morekeywords={RegularPayload, DenimChunk, int32, double, DenimMessage, bytes, bool},
  keywordstyle=\typesCodeStyle,
  classoffset=2, 
  morekeywords={},
  keywordstyle=\builtInCodeStyle,
  classoffset=0,
  morecomment=[s]{/*}{*/},
  morecomment=[l]//,
  morestring=[b]",
  morestring=[b]'
}

\lstdefinelanguage{AnoA}{
  classoffset=0,
  morekeywords={N},
  classoffset=1, 
  morekeywords={if, then, for, else, return},
  keywordstyle=\typesCodeStyle,
  classoffset=2, 
  morekeywords={CR},
  keywordstyle=\builtInCodeStyle,
  classoffset=0,
  morecomment=[s]{/*}{*/},
  morecomment=[l]//,
  morestring=[b]",
  morestring=[b]'
}

\lstdefinelanguage{formalmodel}{
  classoffset=0,
  morekeywords={latest},
  classoffset=1, 
  morekeywords={if, then, for, else, return},
  keywordstyle=\typesCodeStyle,
  classoffset=2, 
  morekeywords={CR},
  keywordstyle=\builtInCodeStyle,
  classoffset=0,
  morecomment=[s]{/*}{*/},
  morecomment=[l]//,
  morestring=[b]",
  morestring=[b]'
}

\crefname{lstlisting}{Listing}{Listings}

\newif\ifanonymized
\anonymizedfalse

\newif\ifverbose
\verbosefalse

\newif\ifdraft
\draftfalse

\newif\ifappendixonly
\appendixonlyfalse

\newif\iffullversion
\fullversiontrue

\newcommand{\ourparagraph}[1]{\textbf{#1.}}

\newcommand{\ourprotocollong}{\textit{Deniable Instant Messaging} (DenIM)\xspace}
\newcommand{\ourprotocol}{DenIM\xspace}
\newcommand{\denimonsignal}{\ourprotocol on Signal\xspace}

\newcommand{\msgsize}{\ensuremath{l}}
\newcommand{\padding}{\ensuremath{q}} 

\newcommand{\attacker}{adversary}

\newcommand{\attackers}{adversaries}

\newcommand{\rsaencryption}{RSAES-OAEP}

\providecommand{\makeCryptoKeyMaterial}[1]{\ensuremath{#1}}
\newcommand{\midtermpublickey}{\makeCryptoKeyMaterial{prepk}}
\newcommand{\midtermprivatekey}{\makeCryptoKeyMaterial{presk}}
\newcommand{\longtermpublickey}{\makeCryptoKeyMaterial{idpk}}
\newcommand{\longtermprivatekey}{\makeCryptoKeyMaterial{idsk}}

\newcommand{\ephemeralpublickey}{\makeCryptoKeyMaterial{epk}}
\newcommand{\ephemeralprivatekey}{\makeCryptoKeyMaterial{esk}}

\newcommand{\messagekey}{\makeCryptoKeyMaterial{mk}}

\newcommand{\ms}{\makeCryptoKeyMaterial{ms}}

\newtheorem{definition}{Definition}
\newtheorem{lemma}{Lemma}
\newtheorem{theorem}{Theorem}

\newcommand{\deniablestyle}[1]{\textcolor{red}{#1}}
\newcommand{\hoststyle}[1]{\textcolor{blue}{#1}}

\newcommand{\eventmeta}{\ensuremath{\eta}}

\newcommand{\countermeta}{\ensuremath{\deniablestyle{c}}}

\newcommand{\keymeta}{\ensuremath{k}}
\newcommand{\keystoremeta}{\ensuremath{{\deniablestyle{K}}}}
\newcommand{\keyothersmeta}{\ensuremath{{\deniablestyle{M}}}}
\newcommand{\keyselfmeta}{\ensuremath{{\deniablestyle{L}}}}

\newcommand{\blockliststoremeta}{\ensuremath{{\deniablestyle{B}}}}
\newcommand{\tokenmeta}{\ensuremath{\mathit{tok}}}
\newcommand{\tokenstoremeta}{\ensuremath{\deniablestyle{R}}}
\newcommand{\seedmeta}{\ensuremath{\deniablestyle{s}}}

\newcommand{\hostpayloadmeta}{\ensuremath{\rho}}
\newcommand{\tracemeta}{\ensuremath{\tau}}
\newcommand\emptytrace{\ensuremath{\epsilon}}
\newcommand{\advnodesmeta}{\ensuremath{\mathbf{N}}}
\newcommand\nodemeta{\ensuremath{n}}

\newcommand\serverstatemeta{\ensuremath{\mathcal{S}}}
\newcommand\serverstatemetaalt{\ensuremath{\mathcal{R}}}
\newcommand\Userstatemeta{\ensuremath{\mathcal{U}}}
\newcommand\Userstatemetaalt{\ensuremath{\mathcal{W}}}
\newcommand\userstatemeta{\ensuremath{u}}
\newcommand\usersignalstatemeta{\ensuremath{\sigma}}

\newcommand{\hoststatemeta}{\ensuremath{\hoststyle{H}}}
\newcommand{\deniablestatemeta}{\ensuremath{\deniablestyle{D}}}
\newcommand{\deniablestatemetaalt}{\ensuremath{\deniablestyle{F}}}

\newcommand\msgmeta{\ensuremath{\alpha}}
\newcommand\msgmetaalt{\ensuremath{\beta}}
\newcommand\dummymeta{\ensuremath{\bullet}}
\newcommand{\messagequeuemeta}{\ensuremath{\mathit{rq}}}
\newcommand{\servereventqueuemeta}{\ensuremath{\mathit{dq}}}

\newcommand{\xaxismeta}{\ensuremath{x}}
\newcommand{\yaxismeta}{\ensuremath{y}}

\newcommand\indexstoremeta{\ensuremath{I}}
\newcommand\firstindexmeta{\ensuremath{w}}
\newcommand\strategymeta{\ensuremath{\omega}}
\newcommand\eventtypemeta{\ensuremath{\kappa}}
\newcommand\keygencounter{\ensuremath{\deniablestyle{g}}}

\newcommand\eventtypesend[1]{\ensuremath{{\mathsf{SEND}\ #1}}}
\newcommand\eventtyperefill{\ensuremath{{\mathsf{REFILL}}}}
\newcommand\eventtypekreq[1]{\ensuremath{{\mathsf{KREQ}\ #1}}}
\newcommand\eventtypeblock[1]{\ensuremath{{\mathsf{BLOCK}\ #1}}}
\newcommand\eventtypedummy{\ensuremath{\bullet}}

\newcommand\keysetmeta{\ensuremath{\mathbf{K}}}

\newcommand\traceprojection[2]{\ensuremath{{\lfloor #1 \rfloor_{#2}}}}

\newcommand{\useridmeta}{\nodemeta}
\newcommand{\regularmsgmeta}{\ensuremath{\gamma}}

\newcommand{\servermessagedownqueue}{\hoststyle{\messagequeuemeta}}

\newcommand{\servereventdownqueue}{\deniablestyle{\servereventqueuemeta}}

\newcommand{\loweq}[3]{\ensuremath{#1 \mathrel{\sim_{#3}} #2}}
\newcommand{\rloweq}[4]{\ensuremath{#1 \mathrel{ {\stackrel{#4}{\sim}}_{#3}}} #2}

\newcommand\initeqmarker{\mathit{init}}
\newcommand{\abstractloweq}[4]{\ensuremath{#1 \mathrel{\simeq_{#3}}^{#4} #2}}

\newcommand{\eventfont}[1]{\mathsf{#1}}
\newcommand{\denimup}[3]{\ensuremath{\eventfont{denimup} (#1, #2, #3)}} 
\newcommand{\denimdown}[4]{\ensuremath{\eventfont{denimdn} (#1, #2, #3, #4)}} 

\newcommand{\evtsend}[3]{\ensuremath{\eventfont{send} (#1 \rightarrow #2, #3)}}
\newcommand{\evtfwd}[3]{\ensuremath{\eventfont{fwd} (#1 \rightarrow #2, #3)}}
\newcommand{\evtblock}[2]{\ensuremath{\eventfont{block} (#1\ \eventfont{by}\ #2)}}
\newcommand{\evtkreq}[2]{\ensuremath{\eventfont{kreq} (#1\ \eventfont{for}\ #2)}}
\newcommand{\evtkresp}[3]{\ensuremath{\eventfont{kresp} ((#1, #2)\ \eventfont{for}\ #3)}}
\newcommand{\evtrefill}[2]{\ensuremath{\eventfont{refill} (#1, #2)}}

\newcommand{\eventdirection}[1]{\ensuremath{\mathrm{dir} ( #1 ) }}
\newcommand{\eventsender}[1]{\ensuremath{\mathrm{snd} ( #1 ) }}
\newcommand{\eventreceiver}[1]{\ensuremath{\mathrm{rcv} ( #1 ) }}
\newcommand{\eventkind}[1]{\ensuremath{\mathrm{kind} ( #1 ) }}
\newcommand{\upstreamevent}{\ensuremath{\uparrow}}
\newcommand{\downstreamevent}{\ensuremath{\downarrow}}

\newcommand{\keytuple}[2]{\ensuremath{(#1, #2)}}
\newcommand{\token}[6]{\ensuremath{\{(#1, #2), (#3, #4), (#5, #6)\}}}
\newcommand{\regulartuple}[2]{\eventfont{msg}(#1, #2)}
\newcommand{\deniablestatetuple}[5]{(#1, #2, #3, #4, #5)} 

\newcommand\systemconfig[3]{\ensuremath{\langle #1, #2, #3 \rangle}}

\newcommand\hostconfig[2]{\ensuremath{\langle #1, #2 \rangle}} 
\newcommand\serverconfig[3]{\ensuremath{\langle #1, #2, #3 \rangle}}

\newcommand\userstratconfig[2]{#1; #2}

\newcommand\networkstep{\ensuremath{\longrightarrow}}
\newcommand\networkstepmany{\ensuremath{\networkstep^{*}}}
\newcommand\serverauxstep[2]{\ensuremath{\underset{#2}{\overset{#1}{\rightrightarrows}}}}
\newcommand\userauxstep[2]{\ensuremath{\underset{}{\overset{#1, #2}{\rightarrowtail}}}}
\newcommand\serverstep[2]{\ensuremath{\xrightarrow[#2]{#1}}}

\newcommand\userdenimstep[2]{\ensuremath{{\userstackrqarrow{#1}{#2}}}}
\newcommand\nouserdenimstep[2]{\ensuremath{{\not\userstackrqarrow{#1}{#2}}}}
\newcommand\userstackrqarrow[2]{%
    \mathrel{\stackunder[2pt]{\stackon[4pt]{$\dashrightarrow$}{$\scriptstyle#1$}}{%
            $\scriptstyle#2$}}}
\newcommand\serverstackrqarrow[2]{%
    \mathrel{\stackunder[2pt]{\stackon[4pt]{$\rightsquigarrow$}{$\scriptstyle#1$}}{%
            $\scriptstyle#2$}}}
\newcommand\serverdenimstep[2]{\ensuremath{{\serverstackrqarrow{#1}{#2}}}}

\newcommand\noserverdenimstep[2]{\ensuremath{{\not \serverstackrqarrow{#1}{#2}}}}
\newcommand\tracecons{\ensuremath{\!\cdot\!}}

\newcommand\userconfig[7]{\ensuremath{\langle #1, #2, #3, #4, #5, #6, #7\rangle}} 

\newcommand{\keyfresh}{\eventfont{fresh}}
\newcommand{\keyused}{\eventfont{used}}

\newcommand{\funprocesshost}[2]{\eventfont{hostresponse} (#1, #2)}

\newcommand{\fundetkeygen}[3]{\eventfont{seededfreshkeys} (#1, #2, #3)} 

\newcommand{\funnext}[2]{\eventfont{next} (#1, #2)}

\newcommand{\dom}[1]{\eventfont{dom}(#1)}
\newcommand{\img}[1]{\eventfont{img}(#1)}

\newcommand\validstrategy[2]{\ensuremath{\mathrm{valid} (#1 \mid #2)}}

\newcommand\noaction{\ensuremath{\bot}}
\newcommand\ruleref[1]{(#1)}
\newcommand\validdests[1]{\mathrm{validdests} (#1)}

\newcommand\formalrandommeta{\ensuremath{\mathsf{r}}}
\newcommand\localrandom[2]{\ensuremath{\mathsf{\$ local} (#1, #2)}}
\newcommand\seededrandom[2]{\ensuremath{\mathsf{\$ seeded} (#1, #2)}}

\newcommand\formalkeyvalue[1]{\ensuremath{\mathsf{k}} \langle #1 \rangle   }

\newcommand\bijectionmeta{\ensuremath{\phi}}

\newcommand\applybijection[2]{#1 (#2)} 
\newcommand\setofformals{\ensuremath{\mathbf{FR}}}
\newcommand\freshorusedmeta{\ensuremath{z}}

\begin{document}


\title{Metadata Privacy Beyond Tunneling for Instant Messaging}

\ifanonymized
\author{{Anonymized submission}}
\else
\author{{Boel Nelson}\\
Aarhus University\\
boel@cs.dk
\and
{Elena Pagnin}\\
Chalmers University of Technology\\
elenap@chalmers.se
\and
{Aslan Askarov}\\
Aarhus University\\
aslan@cs.au.dk
}
\fi

\maketitle


\ifappendixonly
\appendix
\section{Formal model}\label{app:formal-model}

\ifappendixonly
\subsection*{Main theorem (restatement)}
\begin{theorem}[\ourprotocol privacy]\label{thm:denim:privacy}
Consider a set of adversary nodes $\advnodesmeta$, and two initial indistinguishable configurations
$
\abstractloweq{\systemconfig{\serverstatemeta_1}{\Userstatemeta_1}{\emptytrace}}{\systemconfig{\serverstatemeta_2}{\Userstatemeta_2}{\emptytrace}}{\advnodesmeta}{\initeqmarker}
$, with valid user strategies, that is $\validstrategy{\Userstatemeta_i}{\advnodesmeta}, i=1,2$.
If 
$
\systemconfig{\serverstatemeta_1}{\Userstatemeta_1}{\emptytrace} 
\networkstepmany 
\systemconfig{\serverstatemeta'_1}{\Userstatemeta'_1}{\tracemeta_1}
$
then 
$
\systemconfig{\serverstatemeta_2}{\Userstatemeta_2}{\emptytrace} 
\networkstepmany 
\systemconfig{\serverstatemeta'_2}{\Userstatemeta'_2}{\tracemeta_2}
$,
such that $\abstractloweq{\tracemeta_1}{\tracemeta_2}{\advnodesmeta}{}$.
\end{theorem}
\fi

\subsection{Full formal model}\label{app:full-formal-model}

To formally model keys generated in the protocol, we introduce the notion of \emph{formal randomness} that captures two kinds of key generation we have in the protocol. Local key generation takes place on the user nodes when they start new sessions or refill keys. Seeded key generation takes place on the server, and later on the client when their seed counters are updated. We define formal randomness via \emph{tags} defined as follows.
$$
\formalrandommeta ::= \localrandom{\nodemeta}{\keygencounter} \mid
\seededrandom{\seedmeta}{\countermeta}
$$
\noindent 
We denote the set of formal randomness tags as \setofformals.
Keys are represented as values of form $\formalkeyvalue{\formalrandommeta}$, effectively propagating the randomness tag associated with their creation.

\begin{definition}[Regular message]
\begin{align*}
\regularmsgmeta  ::= &\
\regulartuple{\nodemeta}{\hostpayloadmeta}
\end{align*}
\end{definition}

\begin{definition}[User state]
A user state is a tuple \userconfig{\nodemeta}{\keyselfmeta}{\keyothersmeta}{\tokenstoremeta}{\seedmeta}{\countermeta}{\keygencounter}, where
\nodemeta\ is the identity of the user,
\keyselfmeta\ and \keyothersmeta\ are sets containing the user's own keys or keys of other users paired with a state of the form \keytuple{\keymeta}{\keyfresh} or \keytuple{\keymeta}{\keyused},
\tokenstoremeta\ is the abstraction of a ratchets mapped by receiving users to (\firstindexmeta, \{$\keymeta_1, \keymeta_2$\}, $\indexstoremeta$) containing the starting index assigned to the user, the initiating keys, and the observed message indices $\indexstoremeta$,
\seedmeta\ is a seed for the deterministic random number generator,
\countermeta\ is a formal randomness counter for the server-side key generation, and $\keygencounter$ is the formal randomness counter for the client-side key generation.
\end{definition}

\begin{definition}[Server state]
A server state is represented by a tuple 
\serverconfig{\hoststatemeta}{\servermessagedownqueue}{\deniablestatemeta} 
where \hoststatemeta\ represents the state of an arbitrary host protocol, 
\servermessagedownqueue\ is a list of outgoing regular messages, 
and \deniablestatemeta\ represents the deniable state. The deniable state is a mapping of a user to a tuple of the form \deniablestatetuple{\seedmeta}{\countermeta}{\keystoremeta}{\blockliststoremeta}{\servereventdownqueue}, where \seedmeta\ represents a seed for a deterministic random number generator, \countermeta\ a counter, \keystoremeta\ a set of keys, \blockliststoremeta\ a set of blocked users, and \servereventdownqueue\ a list of \ourprotocol\ payloads.
\end{definition}

\subsubsection{Network configuration}
A network configuration is represented by a tuple containing server state, user states and a network trace: \systemconfig{\serverstatemeta}{\Userstatemeta}{\tracemeta}.

\subsubsection{Network and state transitions}
\Cref{fig:network-global-transitions}
\Cref{fig:aux-user-transitions}
\Cref{fig:aux-server-transitions}
\Cref{fig:server-state-transitions}
\Cref{fig:server-state-deniable-processing}
\Cref{fig:state-deniable-upstream-transitions}
\Cref{fig:state-deniable-downstream-transitions}

\begin{figure}
\begin{framed}
\begin{mathpar}
\inferrule[Net-Global]{ 
\Userstatemeta = \userstatemeta_1 \dots \userstatemeta_j \dots \userstatemeta_n\\
{\userstatemeta_j}
\userauxstep{\tracemeta}{\msgmeta}
{\userstatemeta_j}'\\
\Userstatemeta' = \userstatemeta_1 \dots {\userstatemeta_j}' \dots \userstatemeta_n\\
\serverstatemeta
\serverauxstep{}{\msgmeta}
\serverstatemeta'\\
}{ 
\systemconfig{\serverstatemeta}{\Userstatemeta}{\tracemeta}
\networkstep 
\systemconfig{\serverstatemeta'}{\Userstatemeta'}{\tracemeta\tracecons\msgmeta}
}\\
\end{mathpar}
\end{framed}
\caption{Network transitions\label{fig:network-global-transitions}}
\end{figure}

\begin{figure}
\begin{framed}
\begin{mathpar}
\inferrule[User-Denim-Kresp]{
\eventmeta=\evtkresp{\nodemeta_1}{\keymeta}{\nodemeta}\\
\keyothersmeta' = \keyothersmeta [\nodemeta_1 \mapsto \keytuple{\keymeta}{\keyfresh} ]\\
}{
\userconfig{\useridmeta}{\keyselfmeta}{\keyothersmeta}{\tokenstoremeta}{\seedmeta}{\countermeta}{\keygencounter}
\userdenimstep{\eventmeta}{}
\userconfig{\useridmeta}{\keyselfmeta}{\keyothersmeta'}{\tokenstoremeta}{\seedmeta}{\countermeta}{\keygencounter}
}\and 
\inferrule[User-Denim-Fwd-New-Session]{
\eventmeta=\evtfwd{\nodemeta_1}{\nodemeta}{\tokenmeta}\\ 
\nodemeta_1 \notin \dom{\tokenstoremeta}\\
\tokenmeta=\token{\nodemeta_1}{\keymeta_1}{\nodemeta}{\keymeta}{0}{\yaxismeta}\\
\keyselfmeta = \keyselfmeta' \uplus \{\keytuple{\keymeta}{\keyfresh}\}\\
\keyselfmeta'' = \keyselfmeta' \cup \{\keytuple{\keymeta}{\keyused}\}\\
\tokenstoremeta' = \tokenstoremeta[\nodemeta_1 \mapsto (1, \{\keymeta, \keymeta_1\}, \{(0, \yaxismeta)\})]\\
}{
\userconfig{\useridmeta}{\keyselfmeta}{\keyothersmeta}{\tokenstoremeta}{\seedmeta}{\countermeta}{\keygencounter}
\userdenimstep{\eventmeta}{}
\userconfig{\useridmeta}{\keyselfmeta''}{\keyothersmeta}{\tokenstoremeta'}{\seedmeta}{\countermeta}{\keygencounter}
}\\
\inferrule[User-Denim-Fwd-Initialized]{
\eventmeta=\evtfwd{\nodemeta_1}{\nodemeta}{\tokenmeta}\\
\nodemeta_1 \in \dom{\tokenstoremeta}\\
\tokenmeta=\token{\nodemeta_1}{\keymeta_1}{\nodemeta}{\keymeta}{\xaxismeta}{\yaxismeta}\\
\tokenstoremeta(\nodemeta_1) = (\firstindexmeta, \{\keymeta, \keymeta_1\}, \indexstoremeta)\\
\tokenstoremeta' = \tokenstoremeta[\nodemeta_1 \mapsto (\firstindexmeta, \{\keymeta, \keymeta_1\}, \indexstoremeta\cup\{(\xaxismeta, \yaxismeta)\})]\\
}{
\userconfig{\useridmeta}{\keyselfmeta}{\keyothersmeta}{\tokenstoremeta}{\seedmeta}{\countermeta}{\keygencounter}
\userdenimstep{\eventmeta}{}
\userconfig{\useridmeta}{\keyselfmeta}{\keyothersmeta}{\tokenstoremeta'}{\seedmeta}{\countermeta}{\keygencounter}
}
\end{mathpar}
\end{framed}
\caption{Client downstream state transitions\label{fig:state-deniable-downstream-transitions}}
\end{figure}

\begin{figure}
\begin{framed}
\begin{mathpar}
\inferrule[User-Denim-Refill]{
\keymeta =
\formalkeyvalue{\localrandom{\nodemeta}{\keygencounter}} \\
\eventmeta = \evtrefill{\nodemeta}{\keymeta}\\
\keyselfmeta' = \keyselfmeta \cup \{\keytuple{\keymeta}{\keyfresh}\}\\
}{
\userconfig{\useridmeta}{\keyselfmeta}{\keyothersmeta}{\tokenstoremeta}{\seedmeta}{\countermeta}{\keygencounter}
\userdenimstep{}{\eventmeta}
\userconfig{\useridmeta}{\keyselfmeta'}{\keyothersmeta}{\tokenstoremeta}{\seedmeta}{\countermeta}{\keygencounter + 1 }
}\and
\inferrule[User-Denim-Kreq]{
\eventmeta = \evtkreq{\nodemeta_1}{\nodemeta}
}{
\userconfig{\useridmeta}{\keyselfmeta}{\keyothersmeta}{\tokenstoremeta}{\seedmeta}{\countermeta}{\keygencounter}
\userdenimstep{}{\eventmeta}
\userconfig{\useridmeta}{\keyselfmeta}{\keyothersmeta}{\tokenstoremeta}{\seedmeta}{\countermeta}{\keygencounter}
}\and 
\inferrule[User-Denim-Block]{
\eventmeta = \evtblock{\nodemeta}{\nodemeta_1}
}{
\userconfig{\useridmeta}{\keyselfmeta}{\keyothersmeta}{\tokenstoremeta}{\seedmeta}{\countermeta}{\keygencounter}
\userdenimstep{}{\eventmeta}
\userconfig{\useridmeta}{\keyselfmeta}{\keyothersmeta}{\tokenstoremeta}{\seedmeta}{\countermeta}{\keygencounter}
}\and 
\inferrule[User-Denim-Send-New-Session]{
\{\nodemeta, \nodemeta_1\} \notin \dom{\tokenstoremeta}\\
\keymeta = \formalkeyvalue{\localrandom{\nodemeta}{\keygencounter}} \\ 
\keyselfmeta' = \keyselfmeta \cup \{\keytuple{\keymeta}{\keyused}\} \\
\keyothersmeta(\nodemeta_1) = \keytuple{\keymeta_1}{\keyfresh} \\
\keyothersmeta' = \keyothersmeta[\nodemeta_1 \mapsto \keytuple{\keymeta_1}{\keyused}]\\
\tokenmeta = \token{\nodemeta}{\keymeta}{\nodemeta_1}{\keymeta_1}{0}{0}\\
\eventmeta = \evtsend{\nodemeta}{\nodemeta_1}{\tokenmeta}\\
\tokenstoremeta' = \tokenstoremeta[\nodemeta_1\mapsto (0, \{\keymeta, \keymeta_1\}, \{(0,0)\})]\\
}{
\userconfig{\useridmeta}{\keyselfmeta}{\keyothersmeta}{\tokenstoremeta}{\seedmeta}{\countermeta}{\keygencounter}
\userdenimstep{}{\eventmeta}
\userconfig{\useridmeta}{\keyselfmeta'}{\keyothersmeta'}{\tokenstoremeta'}{\seedmeta}{\countermeta}{\keygencounter + 1}
}\and 
\inferrule[User-Denim-Send-Initialized]{
\nodemeta_1 \in \dom{\tokenstoremeta}\\
(\xaxismeta, \yaxismeta) = \funnext{\tokenstoremeta}{\nodemeta_1}\\
\keytuple{\keymeta}{\keyused} \in \keyselfmeta \\
\tokenmeta = \token{\nodemeta}{\keymeta}{\nodemeta_1}{\keymeta_1}{\xaxismeta}{\yaxismeta}\\
\eventmeta = \evtsend{\nodemeta}{\nodemeta_1}{\tokenmeta}\\
\tokenstoremeta(\nodemeta_1) = (\firstindexmeta, \{\keymeta, \keymeta_1\}, \indexstoremeta)\\
\tokenstoremeta' = \tokenstoremeta[\nodemeta \mapsto (\firstindexmeta, \{\keymeta, \keymeta_1\}, \indexstoremeta\cup\{(\xaxismeta, \yaxismeta)\})]\\
}{
\userconfig{\useridmeta}{\keyselfmeta}{\keyothersmeta}{\tokenstoremeta}{\seedmeta}{\countermeta}{\keygencounter}
\userdenimstep{}{\eventmeta}
\userconfig{\useridmeta}{\keyselfmeta}{\keyothersmeta}{\tokenstoremeta'}{\seedmeta}{\countermeta}{\keygencounter}
}
\end{mathpar}
\end{framed}
$$
\begin{array}{l}
\funnext{\tokenstoremeta}{\nodemeta} =  \\ 
\quad \xaxismeta, \yaxismeta = \mathsf{latest}(\tokenstoremeta, \nodemeta) \\ 
\quad \firstindexmeta = \tokenstoremeta(\nodemeta).\firstindexmeta \\ 
\quad \mathsf{if} \xaxismeta \% 2 = \firstindexmeta \\ 
\qquad \mathsf{then}\ (\xaxismeta, \yaxismeta + 1) \\ 
\qquad \mathsf{else}\ (\xaxismeta + 1, 0) \\ 
\end{array}
$$

\caption{Client upstream state transitions, and auxiliary function $\mathsf{next}$, where $\mathit{latest}$ returns the latest index in the ratchet}\label{fig:state-deniable-upstream-transitions}
\end{figure}

\begin{figure}
\begin{framed}
\begin{mathpar}
\inferrule[Aux-Upstream-User-Event]{
\usersignalstatemeta = \userconfig{\useridmeta}{\keyselfmeta}{\keyothersmeta}{\tokenstoremeta}{\seedmeta}{\countermeta}{\keygencounter}\\ 
{\usersignalstatemeta}
\userdenimstep{}{\eventmeta}
{\usersignalstatemeta}'\\
\msgmeta = \denimup{\nodemeta}{\hostpayloadmeta}{\eventmeta} \\
\strategymeta(\traceprojection{\tracemeta}{\nodemeta})  = \eventkind{\eventmeta} 
}{
\userstratconfig{\strategymeta}{\usersignalstatemeta}
\userauxstep{\tracemeta}{\msgmeta}
\userstratconfig{\strategymeta}{\usersignalstatemeta'}\\
}\and 
\inferrule[Aux-Upstream-User-\dummymeta]{
\usersignalstatemeta = \userconfig{\useridmeta}{\keyselfmeta}{\keyothersmeta}{\tokenstoremeta}{\seedmeta}{\countermeta}{\keygencounter}\\ 
\msgmeta = \denimup{\nodemeta}{\hostpayloadmeta}{\dummymeta} \\
\strategymeta(\traceprojection{\tracemeta}{\nodemeta}) = \eventtypedummy
}{ 
\userstratconfig{\strategymeta}{\usersignalstatemeta}
\userauxstep{\tracemeta}{\msgmeta}
\userstratconfig{\strategymeta}{\usersignalstatemeta}\\
}\\
\inferrule[Aux-Downstream-User-Event]{
\usersignalstatemeta = \userconfig{\useridmeta}{\keyselfmeta}{\keyothersmeta}{\tokenstoremeta}{\seedmeta}{\countermeta}{\keygencounter}\\
\msgmeta=\denimdown{\nodemeta}{\hostpayloadmeta}{\eventmeta}{\countermeta'}\\
\keyselfmeta' = \keyselfmeta \cup \fundetkeygen{\seedmeta}{\countermeta}{\countermeta'} \\
\userconfig{\useridmeta}{\keyselfmeta'}{\keyothersmeta}{\tokenstoremeta}{\seedmeta}{\countermeta'}{\keygencounter}
\userdenimstep{\eventmeta}{}
\userconfig{\useridmeta}{\keyselfmeta'}{\keyothersmeta'}{\tokenstoremeta'}{\seedmeta}{\countermeta'}{\keygencounter}\\
\usersignalstatemeta' = \userconfig{\useridmeta}{\keyselfmeta'}{\keyothersmeta'}{\tokenstoremeta'}{\seedmeta}{\countermeta'}{\keygencounter}
}{
\userstratconfig{\strategymeta}{\usersignalstatemeta}
\userauxstep{\tracemeta}{\msgmeta}
\userstratconfig{\strategymeta}{\usersignalstatemeta'}
}\\
\inferrule[Aux-Downstream-User-Dummy-Or-Malformed]{
\usersignalstatemeta = \userconfig{\useridmeta}{\keyselfmeta}{\keyothersmeta}{\tokenstoremeta}{\seedmeta}{\countermeta}{\keygencounter}\\
\msgmeta=\denimdown{\nodemeta}{\hostpayloadmeta}{\eventmeta}{\countermeta'}\\
\keyselfmeta' = \keyselfmeta \cup \fundetkeygen{\seedmeta}{\countermeta}{\countermeta'}\\ 
\usersignalstatemeta' = \userconfig{\useridmeta}{\keyselfmeta'}{\keyothersmeta}{\tokenstoremeta}{\seedmeta}{\countermeta'}{\keygencounter}\\
\usersignalstatemeta' \nouserdenimstep{\eventmeta}{}
}{
\userstratconfig{\strategymeta}{\usersignalstatemeta}
\userauxstep{\tracemeta}{\msgmeta}
\userstratconfig{\strategymeta}{\usersignalstatemeta'}
}
\end{mathpar}
\end{framed}
$$
\fundetkeygen{\seedmeta}{\countermeta}{\countermeta'} = 
    \bigcup_{\countermeta \leq j < \countermeta'} 
       \keytuple{\formalkeyvalue{\seededrandom{\seedmeta}{j}}}{\keyfresh}
$$
\caption{Auxiliary user transitions\label{fig:aux-user-transitions}}
\end{figure}

\begin{figure}
\begin{framed}
\begin{mathpar}
\inferrule[Aux-Upstream-Server-Event]{
\msgmeta = \denimup{\nodemeta}{\hostpayloadmeta}{\eventmeta} \\
\hostconfig{\hoststatemeta}{\servermessagedownqueue}
\serverstep{(\nodemeta, \hostpayloadmeta)}{}
\hostconfig{\hoststatemeta'}{\servermessagedownqueue'}
\\
\deniablestatemeta
\serverdenimstep{\eventmeta}{}
\deniablestatemeta'\\
}{ 
\serverconfig{\hoststatemeta}{\servermessagedownqueue}{\deniablestatemeta}
\serverauxstep{\msgmeta}{}
\serverconfig{\hoststatemeta'}{\servermessagedownqueue'}{\deniablestatemeta'}
}\and 
\inferrule[Aux-Upstream-Server-Dummy-Or-Malformed]{
\msgmeta = \denimup{\nodemeta}{\hostpayloadmeta}{\eventmeta} \\
\hostconfig{\hoststatemeta}{\servermessagedownqueue}
\serverstep{(\nodemeta, \hostpayloadmeta)}{}
\hostconfig{\hoststatemeta'}{\servermessagedownqueue'}
\\
\deniablestatemeta
\noserverdenimstep{\eventmeta}{}
}{ 
\serverconfig{\hoststatemeta}{\servermessagedownqueue}{\deniablestatemeta}
\serverauxstep{\msgmeta}{}
\serverconfig{\hoststatemeta'}{\servermessagedownqueue'}{\deniablestatemeta}
}\\
\inferrule[Aux-Downstream-Server-Event]{
\deniablestatemeta(\nodemeta) = \deniablestatetuple{\seedmeta}{\countermeta}{\keystoremeta}{\blockliststoremeta}{\eventmeta\tracecons\servereventdownqueue}\\
\msgmeta=\denimdown{\nodemeta}{\hostpayloadmeta}{\eventmeta}{\countermeta}\\
\deniablestatemeta' = \deniablestatemeta[\nodemeta_1 \mapsto \deniablestatetuple{\seedmeta}{\countermeta}{\keystoremeta}{\blockliststoremeta}{\servereventdownqueue}]\\
}{
\serverconfig{\hoststatemeta}{(\nodemeta_1, \hostpayloadmeta)\tracecons\servermessagedownqueue}{\deniablestatemeta}
\serverauxstep{\msgmeta}{}
\serverconfig{\hoststatemeta}{\servermessagedownqueue}{\deniablestatemeta'}\\
}\\
\inferrule[Aux-Downstream-Server-\dummymeta]{
\deniablestatemeta(\nodemeta) = \deniablestatetuple{\seedmeta}{\countermeta}{\keystoremeta}{\blockliststoremeta}{[]}\\
\msgmeta=\denimdown{\nodemeta_1}{\hostpayloadmeta}{\dummymeta}{\countermeta}\\
}{
\serverconfig{\hoststatemeta}{(\nodemeta, \hostpayloadmeta)\tracecons\servermessagedownqueue}{\deniablestatemeta}
\serverauxstep{\msgmeta}{}
\serverconfig{\hoststatemeta}{\servermessagedownqueue}{\deniablestatemeta}\\
}\\
\end{mathpar}
\end{framed}
\caption{Auxiliary server transitions\label{fig:aux-server-transitions}}
\end{figure}

\begin{figure}
\begin{framed}
\begin{mathpar}
\inferrule[Server-Host-Action]{
\hoststatemeta', \regularmsgmeta' = \funprocesshost{\hoststatemeta}{\regularmsgmeta}\\
}{ 
\hostconfig{\hoststatemeta}{\servermessagedownqueue}
\serverstep{\regularmsgmeta}{}
\hostconfig{\hoststatemeta'}{\servermessagedownqueue\tracecons\regularmsgmeta'}
}\and 
\inferrule[Server-Host-No-Action]{
\hoststatemeta', \noaction = \funprocesshost{\hoststatemeta}{\regularmsgmeta}\\
}{ 
\hostconfig{\hoststatemeta}{\servermessagedownqueue}
\serverstep{\regularmsgmeta}{}
\hostconfig{\hoststatemeta'}{\servermessagedownqueue}   
}
\end{mathpar}
\end{framed}
\caption{Server state transitions processing host protocol messages\label{fig:server-state-transitions}}
\end{figure}

\begin{figure}
\begin{framed}
\begin{mathpar}
\inferrule[Server-Denim-Refill]{
\eventmeta=\evtrefill{\nodemeta}{\keymeta}\\
\deniablestatemeta(\nodemeta) = \deniablestatetuple{\seedmeta}{\countermeta}{\keystoremeta}{\blockliststoremeta}{\servereventdownqueue}\\
\keystoremeta' = \keystoremeta\cup\keymeta\\
\deniablestatemeta' = \deniablestatemeta[\nodemeta \mapsto \deniablestatetuple{\seedmeta}{\countermeta}{\keystoremeta'}{\blockliststoremeta}{\servereventdownqueue}]\\
}{
{\deniablestatemeta}
\serverdenimstep{\eventmeta}{}
{\deniablestatemeta'}
}\\
\inferrule[Server-Denim-Kreq]{
\eventmeta=\evtkreq{\nodemeta_1}{\nodemeta_2}\\
\deniablestatemeta(\nodemeta_1) =  \deniablestatetuple{\seedmeta_1}{\countermeta_1}{\keystoremeta_1}{\blockliststoremeta_1}{\servereventdownqueue_1} \\
\deniablestatemeta(\nodemeta_2) = \deniablestatetuple{\seedmeta_2}{\countermeta_2}{\keystoremeta_2}{\blockliststoremeta_2}{\servereventdownqueue_2}\\
\keystoremeta_1 = \keystoremeta_1' \uplus \{\keymeta_1\}\\
\eventmeta' = \evtkresp{\nodemeta_1}{\keymeta_1}{\nodemeta_2}\\
\deniablestatemeta' = \deniablestatemeta[\nodemeta_1 \mapsto \deniablestatetuple{\seedmeta_1}{\countermeta_1}{\keystoremeta_1'}{\blockliststoremeta_1}{\servereventdownqueue_1}, \\ \nodemeta_2 \mapsto \deniablestatetuple{\seedmeta_2}{\countermeta_2}{\keystoremeta_2}{\blockliststoremeta_2}{\servereventdownqueue_2\tracecons\eventmeta'}]\\
}{
{\deniablestatemeta}
\serverdenimstep{\eventmeta}{}
\deniablestatemeta'
}\\
\inferrule[Server-Denim-Kreq-Out-of-Keys]{
\eventmeta=\evtkreq{\nodemeta_1}{\nodemeta_2}\\
\deniablestatemeta(\nodemeta_1) = \deniablestatetuple{\seedmeta_1}{\countermeta_1}{\varnothing}{\blockliststoremeta_1}{\servereventdownqueue_1}\\
\deniablestatemeta(\nodemeta_2) = \deniablestatetuple{\seedmeta_2}{\countermeta_2}{\keystoremeta_2}{\blockliststoremeta_2}{\servereventdownqueue_2}\\
\keymeta_1 = \formalkeyvalue{\seededrandom{\seedmeta_1}{\countermeta_1}} \\ %
\eventmeta' = \evtkresp{\nodemeta_1}{\keymeta_1}{\nodemeta_1}\\
\deniablestatemeta' = \deniablestatemeta[\nodemeta_1 \mapsto \deniablestatetuple{\seedmeta_1}{\countermeta_1+1}{\varnothing}{\blockliststoremeta_1}{\servereventdownqueue_1}, \\ \nodemeta_2 \mapsto \deniablestatetuple{\seedmeta_2}{\countermeta_2}{\keystoremeta_2}{\blockliststoremeta_2}{\servereventdownqueue_2\tracecons\eventmeta'}]\\
}{
{\deniablestatemeta}
\serverdenimstep{\eventmeta}{}
{\deniablestatemeta'}
}\\
\inferrule[Server-Denim-Send]{
\eventmeta=\evtsend{\nodemeta_1}{\nodemeta_2}{\tokenmeta}\\
\deniablestatemeta(\nodemeta_2) = \deniablestatetuple{\seedmeta}{\countermeta}{\keystoremeta}{\blockliststoremeta}{\servereventdownqueue}\\
\nodemeta_1 \notin \dom{\blockliststoremeta} \\
\eventmeta' = \evtfwd{\nodemeta_1}{\nodemeta_2}{\tokenmeta} \\
\deniablestatemeta' = \deniablestatemeta[\nodemeta_2 \mapsto \deniablestatetuple{\seedmeta}{\countermeta}{\keystoremeta}{\blockliststoremeta}{\servereventdownqueue\tracecons\eventmeta'}]\\
}{
{\deniablestatemeta}
\serverdenimstep{\eventmeta}{}
{\deniablestatemeta'}
}\\
\inferrule[Server-Denim-Send-Blocked-Or-Unregistered]{
\eventmeta=\evtsend{\nodemeta_1}{\nodemeta_2}{\tokenmeta}\\
\deniablestatemeta(\nodemeta_2) = \deniablestatetuple{\seedmeta}{\countermeta}{\keystoremeta}{\blockliststoremeta}{\servereventdownqueue}\\
(\nodemeta_1 \in \dom{\blockliststoremeta} \lor \nodemeta_2 \notin \dom{\deniablestatemeta})
}{
{\deniablestatemeta}
\serverdenimstep{\eventmeta}{}
{\deniablestatemeta}
}\\
\inferrule[Server-Denim-Block]{
\eventmeta=\evtblock{\nodemeta_2}{\nodemeta_1}\\
\deniablestatemeta(\nodemeta_1) = \deniablestatetuple{\seedmeta}{\countermeta}{\keystoremeta}{\blockliststoremeta}{\servereventdownqueue}\\
\blockliststoremeta' = \blockliststoremeta \cup \{\nodemeta_2\}\\
\deniablestatemeta' = \deniablestatemeta[\nodemeta_1 \mapsto\deniablestatetuple{\seedmeta}{\countermeta}{\keystoremeta}{\blockliststoremeta'}{\servereventdownqueue}]\\
}{
{\deniablestatemeta}
\serverdenimstep{\eventmeta}{}
{\deniablestatemeta'}
}
\end{mathpar}
\end{framed}
\caption{Server state transitions caused by processing of \ourprotocol\ payloads}
\label{fig:server-state-deniable-processing}
\end{figure}

\clearpage

\iffullversion
\subsection{Indistinguishability}\label{app:indistinguishability}
Our low equivalence relations are parameterized by permutation functions 
on formal randomness: $\bijectionmeta: \setofformals \rightarrow \setofformals$. 
The idea is that when comparing configurations from different runs, we replace randomness tags $\formalrandommeta$ with $\bijectionmeta (\formalrandommeta)$. As such, indistinguishability relations use up to two parameters: the set of adversary nodes $\advnodesmeta$ (not relevant for event indistinguishability below) and the permutation function $\bijectionmeta$. We build up indistinguishability bottom-up, starting from events and traces, and leading to system configurations.

\begin{definition}[Event indistinguishability]
We define event indistinguishability structurally as per \Cref{fig:ind:events}.    
\end{definition}

\begin{figure}
\begin{framed}
\begin{mathpar}
\inferrule{~}{
\rloweq{\evtblock{\nodemeta}{\nodemeta}} 
       {\evtblock{\nodemeta}{\nodemeta}}
       {}{\bijectionmeta}
}
\and
\inferrule{~}{
\rloweq{\evtkreq{\nodemeta}{\nodemeta}} 
       {\evtkreq{\nodemeta}{\nodemeta}}
       {}{\bijectionmeta}
}
\and
\inferrule{~}{
\rloweq{\dummymeta} 
       {\dummymeta}
       {}{\bijectionmeta}
}\and
\inferrule{
    \tokenmeta_1 = 
       \token{\nodemeta_1}
             {\formalkeyvalue{\formalrandommeta_1}}
             {\nodemeta_2}
             {\formalkeyvalue{\formalrandommeta_2}}
             {\xaxismeta}
             {\yaxismeta}
            \\ 
    \tokenmeta_2 = 
       \token{\nodemeta_1}
             {\formalkeyvalue{\bijectionmeta (\formalrandommeta_1)}}
             {\nodemeta_2}
             {\formalkeyvalue{\bijectionmeta (\formalrandommeta_2)}}
             {\xaxismeta}
             {\yaxismeta}
}{
\rloweq{\evtsend{\nodemeta_1}{\nodemeta_2}{\tokenmeta_1}} 
       {\evtsend{\nodemeta_1}{\nodemeta_2}{\tokenmeta_2}}
       {}{\bijectionmeta}
}
\and
\inferrule{
    \tokenmeta_1 = 
       \token{\nodemeta_1}
             {\formalkeyvalue{\formalrandommeta_1}}
             {\nodemeta_2}
             {\formalkeyvalue{\formalrandommeta_2}}
             {\xaxismeta}
             {\yaxismeta}
            \\ 
    \tokenmeta_2 = 
       \token{\nodemeta_1}
             {\formalkeyvalue{\bijectionmeta (\formalrandommeta_1)}}
             {\nodemeta_2}
             {\formalkeyvalue{\bijectionmeta (\formalrandommeta_2)}}
             {\xaxismeta}
             {\yaxismeta}
}{
\rloweq{\evtfwd{\nodemeta_1}{\nodemeta_2}{\tokenmeta_1}} 
       {\evtfwd{\nodemeta_1}{\nodemeta_2}{\tokenmeta_2}}
       {}{\bijectionmeta}
}
\and
\inferrule{~}{
\rloweq{\evtrefill{\nodemeta}{\formalkeyvalue{\formalrandommeta}}} 
       {\evtrefill{\nodemeta}{\formalkeyvalue{\bijectionmeta (\formalrandommeta)}}}
       {}{\bijectionmeta}
}
\and
\inferrule{~}{
\rloweq{\evtkresp{\nodemeta_1}{\formalkeyvalue{\formalrandommeta}}{\nodemeta_2}}
       {\evtkresp{\nodemeta_1}{\formalkeyvalue{\bijectionmeta (\formalrandommeta)}}{\nodemeta_2}}
       {}{\bijectionmeta}
}
\end{mathpar}
\end{framed}
\caption{Event indistinguishability\label{fig:ind:events}}
\end{figure}

\begin{definition}[Message indistinguishability]
Given two messages $\msgmeta_1$ and $\msgmeta_2$, define \emph{message indistinguishability} w.r.t. a set of adversarial nodes $\advnodesmeta$, and permutation $\bijectionmeta$, written $\rloweq{\msgmeta_1}{\msgmeta_2}{\advnodesmeta}{\bijectionmeta}$, 
based on the structure of the messages, as follows:

\begin{mathpar}
\inferrule{
\nodemeta \in \advnodesmeta \implies 
    \rloweq{\eventmeta_1}{\eventmeta_2}{}{\bijectionmeta}
}{
\rloweq{
  \denimup{\nodemeta}{\hostpayloadmeta}{\eventmeta_2}}{
  \denimup{\nodemeta}{\hostpayloadmeta}{\eventmeta_1}}{
  \advnodesmeta}{
  \bijectionmeta
  }
}
\and 
\inferrule{
\nodemeta \in \advnodesmeta \implies 
    \rloweq{\eventmeta_1}{\eventmeta_2}{}{\bijectionmeta} 
    \land \countermeta_1 = \countermeta_2
}{
\rloweq{
  \denimdown{\nodemeta}{\hostpayloadmeta}{\eventmeta_1}{\countermeta_1}}{
  \denimdown{\nodemeta}{\hostpayloadmeta}{\eventmeta_1}{\countermeta_1}}{
  \advnodesmeta}{
  \bijectionmeta
  }
}
\end{mathpar}

\end{definition}

\begin{definition}[Trace indistinguishability]
Consider two traces $\tracemeta_1$, $\tracemeta_2$ of length $n$, each composed of events $\msgmeta_{1i}$ and $ \msgmeta_{2i}$ respectively. Consider  a set of adversarial 
nodes $\advnodesmeta$. Say that two traces are \emph{indistinguishable}
to the set of attacker nodes $\advnodesmeta$, written
\loweq{\tracemeta_1}{\tracemeta_2}{\advnodesmeta},
if for all $i = 1.. n$, it holds that $\rloweq{\msgmeta_{1i}}{\msgmeta_{2i}}{\advnodesmeta}{\bijectionmeta}$.
\end{definition}

We use notation $\abstractloweq{\tracemeta_1}{\tracemeta_2}{\advnodesmeta}{\initeqmarker}$ if 
$\rloweq{\tracemeta_1}{\tracemeta_2}{\advnodesmeta}
{\bijectionmeta_{\initeqmarker}}$
holds for the empty bijection $\bijectionmeta_{\initeqmarker}$. We write 
$\abstractloweq{\tracemeta_1}{\tracemeta_2}{\advnodesmeta}{}$, if there exists 
a bijection $\bijectionmeta$ such that 
$\rloweq{\tracemeta_1}{\tracemeta_2}{\advnodesmeta}{\bijectionmeta}$.

\begin{definition}[User state indistinguishability]
\begin{mathpar}
\inferrule{
\usersignalstatemeta_1 = 
    \userconfig{\useridmeta}{\keyselfmeta_1}{\keyothersmeta_2}{\tokenstoremeta_1}{\seedmeta}{\countermeta}{\keygencounter} \\
\usersignalstatemeta_2 = \userconfig{\useridmeta}{\keyselfmeta_2}{\keyothersmeta_2}{\tokenstoremeta_2}{\seedmeta}{\countermeta}{\keygencounter} \\
\useridmeta \in \advnodesmeta\\
\keytuple{\formalkeyvalue{\formalrandommeta}}{\freshorusedmeta} \in \keyselfmeta_1 \Leftrightarrow 
\keytuple{\formalkeyvalue{\applybijection \bijectionmeta \formalrandommeta}}{\freshorusedmeta} \in \keyselfmeta_2 \\
\keyothersmeta(\nodemeta') =
   \keytuple{\formalkeyvalue{\formalrandommeta}}{\freshorusedmeta} 
\Leftrightarrow
\keyothersmeta(\nodemeta') =
    \keytuple{\formalkeyvalue{\applybijection \bijectionmeta \formalrandommeta}}{\freshorusedmeta}   
\\
\tokenstoremeta(\nodemeta') =
     (\firstindexmeta, 
     \{ \formalkeyvalue{\formalrandommeta} , 
        \formalkeyvalue{\formalrandommeta_1} \}, 
     \indexstoremeta)
\Leftrightarrow
\tokenstoremeta(\nodemeta') =
     (\firstindexmeta, 
     \{ \formalkeyvalue{\applybijection \bijectionmeta \formalrandommeta}, 
        \formalkeyvalue{\applybijection \bijectionmeta {\formalrandommeta_1}}
      \}, 
     \indexstoremeta)
}{
  \rloweq{
  \userstratconfig{\strategymeta}{\usersignalstatemeta_1}
   }
  {
  \userstratconfig{\strategymeta}{\usersignalstatemeta_2}
  }{\advnodesmeta}
  {\bijectionmeta}
}\\
\inferrule{
\usersignalstatemeta_1  = \userconfig{\useridmeta}{\keyselfmeta_1}{\keyothersmeta_1}{\tokenstoremeta_1}{\seedmeta_1}{\countermeta_1}{\keygencounter_1} \and 
\usersignalstatemeta_2 = \userconfig{\useridmeta}{\keyselfmeta_2}{\keyothersmeta_2}{\tokenstoremeta_2}{\seedmeta_2}{\countermeta_2}{\keygencounter_2} \\ 
\useridmeta \notin \advnodesmeta 
}{
\rloweq{ \userstratconfig{\strategymeta_1}{\usersignalstatemeta_1}
}{
\userstratconfig{\strategymeta_2}{\usersignalstatemeta_2}
}{\advnodesmeta}
{\bijectionmeta}
}

\end{mathpar}
\end{definition}

\begin{definition}[Server user configuration indistinguishability]
\begin{mathpar}
\inferrule{
\formalkeyvalue{\formalrandommeta} \in \keystoremeta_1 
\Leftrightarrow
\formalkeyvalue{\applybijection \bijectionmeta \formalrandommeta} \in \keystoremeta_2 
\\
\servereventdownqueue_1= \eventmeta^1_1  \dots \eventmeta^1_j \\
\servereventdownqueue_2= \eventmeta^2_1  \dots \eventmeta^2_j \\
\rloweq{\eventmeta^1_i}{\eventmeta^2_i}{}{\bijectionmeta}, \quad i = 1..j
}{
\rloweq{
    \deniablestatetuple
        {\seedmeta}{\countermeta}{\keystoremeta_1}{\blockliststoremeta}{\servereventdownqueue_1}}
{\deniablestatetuple{\seedmeta}{\countermeta}{\keystoremeta_2}{\blockliststoremeta}{\servereventdownqueue_2}} 
{}
{\bijectionmeta}
}
\end{mathpar}
\end{definition}

\begin{definition}[System configuration indistinguishability]
Given two system configurations $\systemconfig{\serverstatemeta_1}{\Userstatemeta_1}{\tracemeta_1}$, $\systemconfig{\serverstatemeta_2}{\Userstatemeta_2}{\tracemeta_2}$, we say that two configurations are \emph{indistinguishable}
written
\loweq{\systemconfig{\serverstatemeta_1}{\Userstatemeta_1}{\tracemeta_1}}{\systemconfig{\serverstatemeta_2}{\Userstatemeta_2}{\tracemeta_2}}{\advnodesmeta},
if they are indistinguishable component-wise. Technically, 

\begin{mathpar}
\inferrule{
\serverstatemeta_1 = 
\serverconfig{\hoststatemeta}{\servermessagedownqueue}{\deniablestatemeta_1} \and 
\serverstatemeta_2 = \serverconfig{\hoststatemeta}{\servermessagedownqueue}{\deniablestatemeta_2} \\
\forall \nodemeta \in \advnodesmeta~.~
    \rloweq{    \deniablestatemeta_1  (\nodemeta)}{ \deniablestatemeta_2 (\nodemeta)}{}{\bijectionmeta} \\
\Userstatemeta_1 = \usersignalstatemeta^1_{1} \dots \usersignalstatemeta^1_{j} \dots \usersignalstatemeta^1_{k} \and  
\Userstatemeta_2 = \usersignalstatemeta^2_{1} \dots \usersignalstatemeta^2_{j} \dots \usersignalstatemeta^2_{k} \\ 
\rloweq{\usersignalstatemeta^1_j}{\usersignalstatemeta^2_j}{\advnodesmeta}{\bijectionmeta}, j = 1..k \and 
\rloweq{\tracemeta_1}{\tracemeta_2}{\advnodesmeta}{\bijectionmeta}
 \\
}{
\rloweq{\systemconfig{\serverstatemeta_1}{\Userstatemeta_1}{\tracemeta_1}}{\systemconfig{\serverstatemeta_2}{\Userstatemeta_2}{\tracemeta_2}}{\advnodesmeta}{\bijectionmeta}
}
\end{mathpar}
\end{definition}

\subsection{Proof of \Cref{thm:denim:privacy}}\label{app:main-proof}

\begin{definition}[Strategy determinism  w.r.t. formal randomness]
A strategy is \emph{deterministic} w.r.t formal randomness, if for any two 
traces $\tracemeta_1$ and $\tracemeta_2$, and any $\advnodesmeta, \bijectionmeta$ 
such that $\rloweq{\tracemeta_1}{\tracemeta_2 }{\advnodesmeta}{\bijectionmeta}$, 
it holds that $\strategymeta (\tracemeta_1) = \strategymeta (\tracemeta_2)$.
\end{definition}

\begin{definition}[Valid destinations]
Given a signal configuration $\usersignalstatemeta =
\userconfig{\useridmeta}{\keyselfmeta}{\keyothersmeta}{\tokenstoremeta}{\seedmeta}{\countermeta}{\keygencounter}$,
define \emph{valid destinations} for this configuration, denoted as $\validdests{\usersignalstatemeta}$, as 
$\validdests{\usersignalstatemeta} = \dom\keyothersmeta \cup \dom \tokenstoremeta$.
\end{definition}

\begin{definition}[Tight bijections]
Given a pair of configurations 
$\systemconfig{\serverstatemeta}{\Userstatemeta}{\tracemeta_1}$ and $\systemconfig{\serverstatemetaalt}{\Userstatemetaalt}{\tracemeta_2}$, and a set of adversarial nodes $\advnodesmeta$, say 
that a bijection $\bijectionmeta$ is \emph{tight} w.r.t. these configurations and 
$\advnodesmeta$, if $\dom{ \bijectionmeta }$ is restricted to formal randomness tags that appear in the adversarial components of $\serverstatemeta$ and $\Userstatemeta$, and $\img {\bijectionmeta}$ is restricted to formal randomness tags that appear in the the adversarial components of $\serverstatemetaalt$ and $\Userstatemetaalt$.
\end{definition}

\begin{lemma}[Unwinding]\label{lemma:unwinding}
Suppose a set of adversarial nodes $\advnodesmeta$ and 
two configurations $\systemconfig{\serverstatemeta}{\Userstatemeta}{\tracemeta_1}$ and $\systemconfig{\serverstatemetaalt}{\Userstatemetaalt}{\tracemeta_2}$, such that 
\begin{itemize}[noitemsep,nolistsep]
\item there is a tight bijection $\bijectionmeta$ such that
    $\rloweq{\systemconfig{\serverstatemeta}{\Userstatemeta}{\tracemeta_1}}
           {\systemconfig{\serverstatemetaalt}{\Userstatemetaalt}{\tracemeta_2}}
           {\advnodesmeta}
           {\bijectionmeta}
           $, and 
\item all user strategies are valid, i.e., $\validstrategy{\Userstatemeta}{\advnodesmeta}$
and $\validstrategy{\Userstatemetaalt}{\advnodesmeta}$
\item traces $\tracemeta_1$ (resp. $\tracemeta_2$) are consistent with valid destinations for user signal configurations in $\Userstatemeta$ (resp. $\Userstatemetaalt)$, i.e.,
      for any pair signal state ${\usersignalstatemeta}$ in $\Userstatemeta$ (resp. $\Userstatemetaalt) $ of node $\nodemeta$, if for any valid strategy $\strategymeta$ it holds that 
     $\strategymeta (\traceprojection{\tracemeta}{\nodemeta}) = \eventtypesend{\nodemeta_{\mathit{dest}}}$, then 
     $\nodemeta_{\mathit{dest}} \in \validdests{\usersignalstatemeta}$

\item 
there is a message $\msgmeta$ that is valid w.r.t. $\advnodesmeta$ such that
$
\systemconfig{\serverstatemeta}{\Userstatemeta}{\tracemeta_1} \networkstep  
\systemconfig{\serverstatemeta'}{\Userstatemeta'}{\tracemeta_1 \tracecons \msgmeta} 
$
\end{itemize}
Then there is a message $\msgmetaalt$, and a tight bijection $\bijectionmeta'$ that 
extends $\bijectionmeta$, such that
$
\systemconfig{\serverstatemetaalt}{\Userstatemetaalt}{\tracemeta_2} \networkstep  
\systemconfig{\serverstatemetaalt'}{\Userstatemetaalt'}{\tracemeta_2 \tracecons \msgmetaalt} 
$
and 
    $\rloweq{\systemconfig{\serverstatemeta'}{\Userstatemeta'}{\tracemeta_1 \tracecons  \msgmeta }}
           {\systemconfig{\serverstatemetaalt'}{\Userstatemetaalt'}{\tracemeta_2  \tracecons \msgmetaalt }}
           {\advnodesmeta}{\bijectionmeta'}$
and extended traces $\tracemeta_1 \tracecons  \msgmeta$ (resp. $\tracemeta_2  \tracecons \msgmetaalt $) are  consistent with valid destinations in the updated configurations $\Userstatemeta$ (resp. $\Userstatemetaalt$).

\end{lemma}
\begin{proof}
Let $\serverstatemeta = \serverconfig{\hoststatemeta}{\servermessagedownqueue}{\deniablestatemeta}$ and 
$\serverstatemetaalt = \serverconfig{\hoststatemeta}{\servermessagedownqueue}{\deniablestatemetaalt}$. 
The only way to produce the message $\msgmeta$ at the network level
is by rule \ruleref{Net-Global}.
By inversion of the rule, there is a user state $\userstatemeta_j$ for which it holds that 
$\userstatemeta_j \userauxstep{\tracemeta_1}{\msgmeta} \userstatemeta'_j$
and 
$\serverstatemeta \serverauxstep{\msgmeta}{} \serverstatemeta'$.
We proceed by case analysis on $\msgmeta$.
\begin{description}[noitemsep,nolistsep]
\item[Case $\msgmeta = \denimup{n}{\hostpayloadmeta}{\eventmeta_1}$]

On the server side, the two possible transitions are \ruleref{Aux-Upstream-Server-Event} or \ruleref{Aux-Upstream-Server-Dummy-Or-Malformed}. In either case, the host server is updated through the host response function 
$\hoststatemeta',  \regularmsgmeta = \funprocesshost{\hoststatemeta}{(\nodemeta, \hostpayloadmeta)}$, where $\regularmsgmeta$ is a potential host response, and the reply
queue is potentially extended depending on $\regularmsgmeta$, per rules \ruleref{Server-Host-Action} and \ruleref{Server-Host-No-Action}.
We proceed by distinguishing whether the sending node is adversarial or not. 

\begin{description}[noitemsep,nolistsep]
\item [Sending node is not part of $\advnodesmeta$]
In this case, the adversary can observe only the host protocol aspects of the message. 

We consider two sub-cases, depending on whether there is a transition
$\deniablestatemeta \serverdenimstep{\eventmeta_1}{} \deniablestatemeta'$. 
\begin{description}[noitemsep,nolistsep]
    \item[Case $\deniablestatemeta \serverdenimstep{\eventmeta_1}{} \deniablestatemeta'$]
    Because $\msgmeta$ is produced by a valid non-adversarial strategy, it does not contain messages that modify the parts of the server state for nodes in $\advnodesmeta$. 

    \item[Case $\deniablestatemeta \noserverdenimstep{\eventmeta_1}{}$]. We hit this case, if $\eventmeta_2$ is dummy $\dummymeta$, or it is a malformed event and no premises of \Cref{fig:server-state-deniable-processing} are satisfied. The state of the nodes in $\advnodesmeta$ is not affected.    
\end{description}

\vspace{0.5em}

Suppose now that $\strategymeta_2$ is the strategy of user $\nodemeta$ in configuration $\Userstatemetaalt$. 
Let $\strategymeta_2 (\tracemeta_2) = \eventtypemeta_2 $ be the decision of the strategy at this point. Because
the strategy is valid it must be the case that we can construct a matching DenIM event $\eventmeta_2$ such that
$\eventkind{\eventmeta_2} = \eventtypemeta_2$.

\vspace{0.5em}
Let $\msgmetaalt = \denimup{n}{\hostpayloadmeta}{\eventmeta_2}$. Similar to the above reasoning about $\msgmeta$, we can argue that the state of the adversarial nodes on the server is not affected by this message, regardless of whether the server uses the rules where $\deniablestatemetaalt \serverdenimstep{\eventmeta_2}{} \deniablestatemetaalt'$ or $\deniablestatemetaalt \noserverdenimstep{\eventmeta_2}{}$.

\vspace{0.5em}
Because $\msgmeta$ is an upstream message, it does not change the set of valid destination for user $\nodemeta$. For the same reason, if a valid strategy chooses a destination based on the trace $\tracemeta_1 \tracecons \msgmeta$, the same
destination is allowed to be chosen for the trace $\tracemeta_1$. This means that the extended trace is consistent with respect to the updated signal configuration for user $\nodemeta$ (all other users are unchanged). 
Similar argument holds for $\msgmetaalt$.

\vspace{0.5em}
Finally, because no attacker-visible keys are generated by these messages, we keep the bijection, and let $\bijectionmeta' = \bijectionmeta$.
\vspace{0.5em}

Putting everything together we have that both the server and the user states are extended in a way that is not distinguishable to the adversary; and that the updated traces are consistent with valid destinations, which concludes this case of the proof.

\item [Sending node is part of $\advnodesmeta$]
We consider two sub-cases, depending on whether there is a transition
$\deniablestatemeta \serverdenimstep{\eventmeta_1}{} \deniablestatemeta'$. 

\begin{description}[noitemsep,nolistsep]
    \item[Case $\deniablestatemeta \noserverdenimstep{\eventmeta_1}{}$]. We hit this case, if $\eventmeta_2$ is dummy $\dummymeta$, or it is a malformed event and no premises of \Cref{fig:server-state-deniable-processing} are satisfied. The state of the nodes in $\advnodesmeta$ is not affected.    

    \item[Case $\deniablestatemeta \serverdenimstep{\eventmeta_1}{} \deniablestatemeta'$]
   This message modifies the server state corresponding to the adversary. However, because this is an adversarial node, it means that configurations $\Userstatemeta$ and $\Userstatemetaalt$ agree on both the signal state and the strategy $\strategymeta$ used by this node. 
   We have that $\eventtypemeta_1 = \strategymeta(\traceprojection{\tracemeta_1}{\nodemeta})$ and $\eventtypemeta_2 = \strategymeta(\traceprojection{\tracemeta_2}{\nodemeta})$. Because 
   the traces are indistinguishable and the strategy is valid, 
   $\eventtypemeta_1 = \eventtypemeta_2$.
   We proceed by considering the different non-dummy possibilities for $\eventtypemeta_1$.
   \begin{description}[noitemsep,nolistsep]
        \item [Case \eventtypesend{\nodemeta'} ] If this message initiates a new session, cf. rule \ruleref{User-Denim-Send-New-Session}, a fresh key is generated. Because the two configurations agree on the key generation counters, the keys generated in both runs can be matched in the extension  of the bijection. If no new keys are generated at this step, which means that sending happens along an established session cf \ruleref{User-Denim-Send-Initialized}, then the bijection remains unchanged. 
        \item [Case  \eventtyperefill ] Similar to the above, we can extend the bijection in both runs.
        \item [Case \eventtypekreq{\nodemeta'}] There are two ways the server can process this message 
                \begin{enumerate}[noitemsep,nolistsep]
                    \item There are available keys for $\nodemeta'$ on the server, cf. rule \ruleref{Server-Denim-Kreq}, and the server picks one of the keys from the keystore of $\nodemeta'$. We know that these keys are not in the domain/image of the bijection, because the bijection is tight.
                    \item There are no available keys for $\nodemeta'$ on the server, cf. rule \ruleref{Server-Denim-Kreq-Out-of-Keys}, and the server generates a seeded key. These keys are also not in the domain/image of the bijection. 
                \end{enumerate}
             Let $\formalkeyvalue{\formalrandommeta_1}$ be the key obtained in the first run, and $\formalkeyvalue{\formalrandommeta_2}$ be the key generated in the second run, and we know that $\formalrandommeta_1$ and $\formalrandommeta_1$ are not in the bijection.  We can therefore extend $\bijectionmeta$ by mapping $\formalrandommeta_1$ to $\formalrandommeta_2$. This extension preserves tightness property of the bijection, because by moving them to the deniable output queue on the server, they are now part of the adversarial state.
                    
        \item [Case \eventtypeblock{\nodemeta'}] This case does not affect the generated keys; and therefore can be matched exactly in both runs, without extending the bijection.
   \end{description}
To conclude this case, we observe that when the adversary state on the server is extended, it happens in a way that can be matched in the second run, including matching new keys, which concludes the case.
  
\end{description}

\end{description}

\item[Case $\msgmeta = \denimdown{n}{\hostpayloadmeta}{\eventmeta_1}{\countermeta_1}$] We consider two cases.
\begin{description}
\item [The receiving node is not part of $\advnodesmeta$] This does not change any parts of the adversary state, and we are done immediately.
\item [The receiving node is part of $\advnodesmeta$] The parts of the adversarial state are changed in the way that is matched across both runs, but no new keys are generated, and we therefore are done immediately as well.  
\end{description}

\end{description}
\end{proof}

Using the \Cref{lemma:unwinding}, the proof of \Cref{thm:denim:privacy} is immediate by induction on the trace $\tracemeta_1$.

\fi
\section{Experiment plots}\label{app:experiment-graphs}
\FloatBarrier

\begin{figure}[htb]
     \centering
     \begin{subfigure}[t]{0.3\textwidth}
         \centering
         \includegraphics[width=\textwidth]{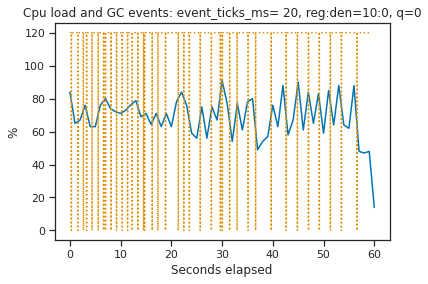}
         \caption{CPU load over time.}
         \label{fig:graph-cpu-a1}
     \end{subfigure}
     \hfill
     \begin{subfigure}[t]{0.31\textwidth}
         \centering
         \includegraphics[width=\textwidth]{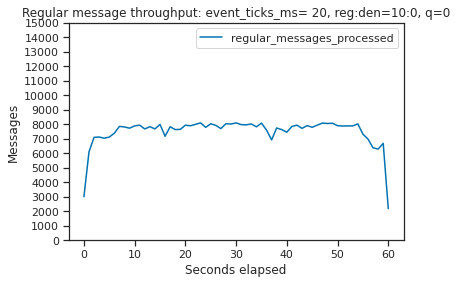}
         \caption{Regular messages processed over time.}
         \label{fig:graph-throughput-a1}
     \end{subfigure}
     \begin{subfigure}[t]{0.37\textwidth}
         \centering
         \includegraphics[width=\textwidth]{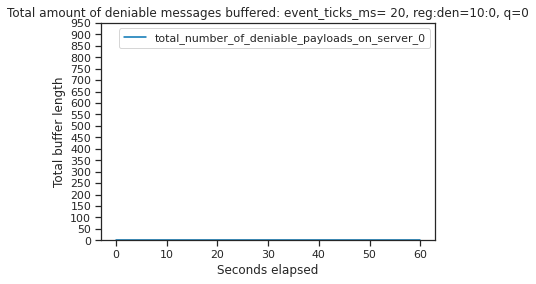}
         \caption{Deniable buffer length over time. Notice that since $\padding=0$, no deniable traffic is piggybacked to the server.}
         \label{fig:graph-buffer-a1}
     \end{subfigure}
\caption{Server statistics collected with setting A1.}\label{fig:graph-server-statistics-a1}
\end{figure}
 
\begin{figure}[htb]
     \centering
     \begin{subfigure}[b]{0.3\textwidth}
         \centering
         \includegraphics[width=\textwidth]{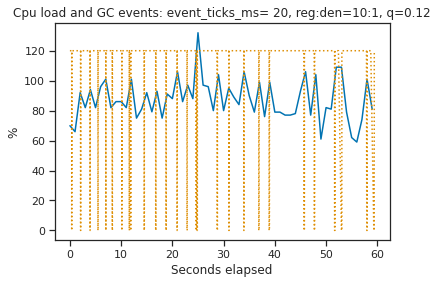}
         \caption{CPU load over time.}
         \label{fig:graph-cpu-a2}
     \end{subfigure}
     \hfill
     \begin{subfigure}[b]{0.31\textwidth}
         \centering
         \includegraphics[width=\textwidth]{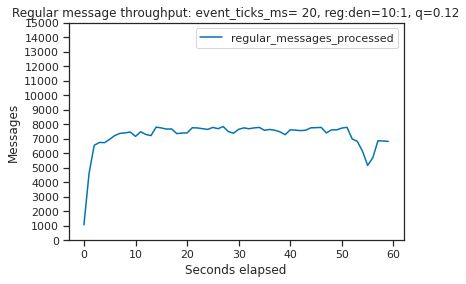}
         \caption{Regular messages processed over time.}
         \label{fig:graph-throughput-a2}
     \end{subfigure}
     \begin{subfigure}[b]{0.37\textwidth}
         \centering
         \includegraphics[width=\textwidth]{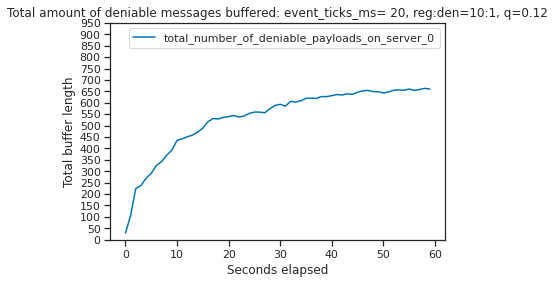}
         \caption{Deniable buffer length over time.}
         \label{fig:graph-buffer-a2}
     \end{subfigure}
\caption{Server statistics collected with setting A2.}\label{fig:graph-server-statistics-a2}
\end{figure}

\begin{figure}[htb]
     \centering
     \begin{subfigure}[b]{0.3\textwidth}
         \centering
         \includegraphics[width=\textwidth]{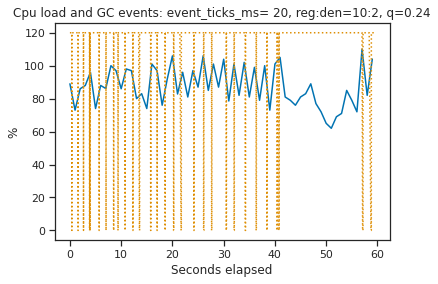}
         \caption{CPU load over time.}
         \label{fig:graph-cpu-a3}
     \end{subfigure}
     \hfill
     \begin{subfigure}[b]{0.31\textwidth}
         \centering
         \includegraphics[width=\textwidth]{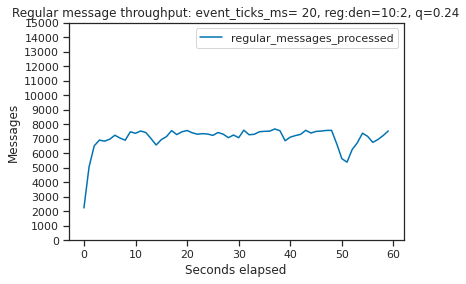}
         \caption{Regular messages processed over time..}
         \label{fig:graph-throughput-a3}
     \end{subfigure}
     \begin{subfigure}[b]{0.37\textwidth}
         \centering
         \includegraphics[width=\textwidth]{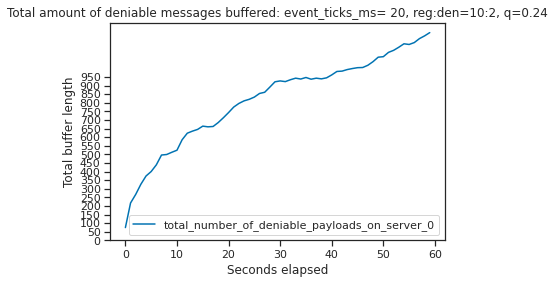}
         \caption{Deniable buffer length over time}
         \label{fig:graph-buffer-a3}
     \end{subfigure}
\caption{Server statistics collected with setting A3.}\label{fig:graph-server-statistics-a3}
\end{figure}

\begin{figure}[htb]
     \centering
     \begin{subfigure}[b]{0.3\textwidth}
         \centering
         \includegraphics[width=\textwidth]{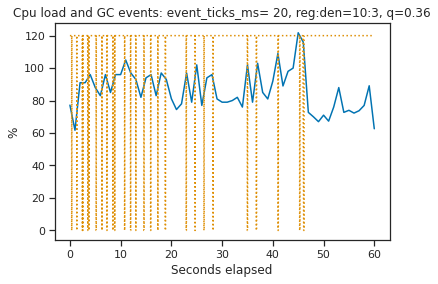}
         \caption{CPU load over time.}
         \label{fig:graph-cpu-a4}
     \end{subfigure}
     \hfill
     \begin{subfigure}[b]{0.31\textwidth}
         \centering
         \includegraphics[width=\textwidth]{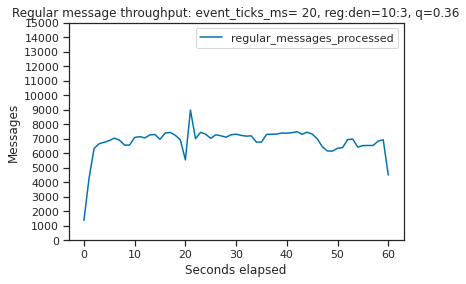}
         \caption{Regular messages processed over time.}
         \label{fig:graph-throughput-a4}
     \end{subfigure}
     \begin{subfigure}[b]{0.37\textwidth}
         \centering
         \includegraphics[width=\textwidth]{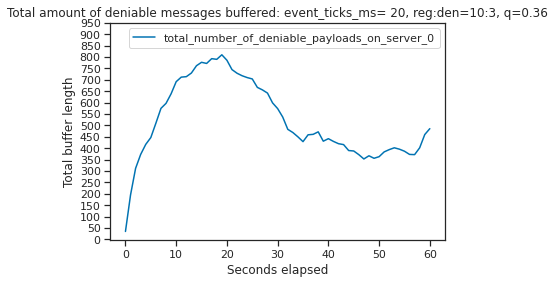}
         \caption{Deniable buffer length over time.}
         \label{fig:graph-buffer-a4}
     \end{subfigure}
\caption{Server statistics collected with setting A4.}\label{fig:graph-server-statistics-a4}
\end{figure}

\begin{figure}[htb]
     \centering
     \begin{subfigure}[b]{0.3\textwidth}
         \centering
         \includegraphics[width=\textwidth]{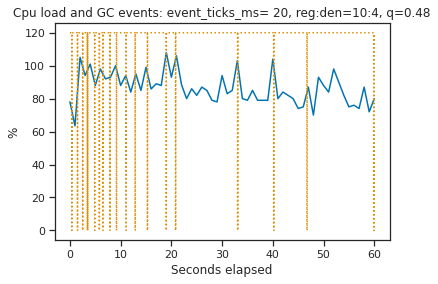}
         \caption{CPU load over time.}
         \label{fig:graph-cpu-a5}
     \end{subfigure}
     \hfill
     \begin{subfigure}[b]{0.31\textwidth}
         \centering
         \includegraphics[width=\textwidth]{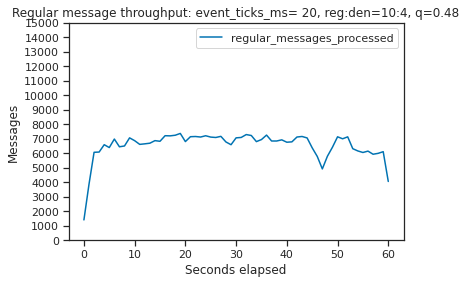}
         \caption{Regular messages processed over time.}
         \label{fig:graph-throughput-a5}
     \end{subfigure}
     \begin{subfigure}[b]{0.37\textwidth}
         \centering
         \includegraphics[width=\textwidth]{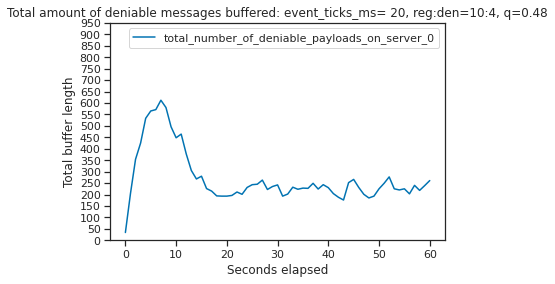}
         \caption{Deniable buffer length over time.}
         \label{fig:graph-buffer-a5}
     \end{subfigure}
\caption{Server statistics collected with setting A5.}\label{fig:graph-server-statistics-a5}
\end{figure}

\begin{figure}[htb]
     \centering
     \begin{subfigure}[b]{0.3\textwidth}
         \centering
         \includegraphics[width=\textwidth]{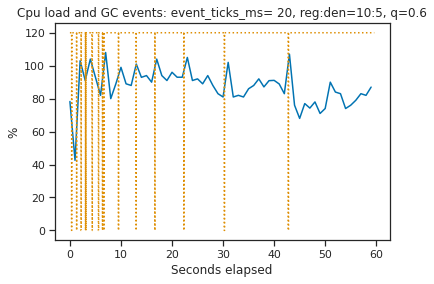}
         \caption{CPU load over time.}
         \label{fig:graph-cpu-a6}
     \end{subfigure}
     \hfill
     \begin{subfigure}[b]{0.31\textwidth}
         \centering
         \includegraphics[width=\textwidth]{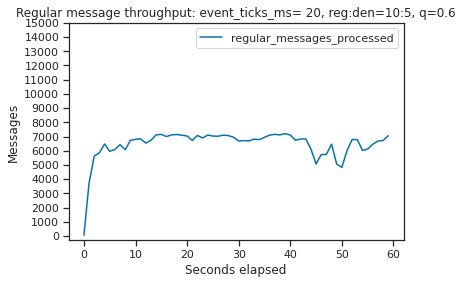}
         \caption{Regular messages processed over time.}
         \label{fig:graph-throughput-a6}
     \end{subfigure}
     \begin{subfigure}[b]{0.37\textwidth}
         \centering
         \includegraphics[width=\textwidth]{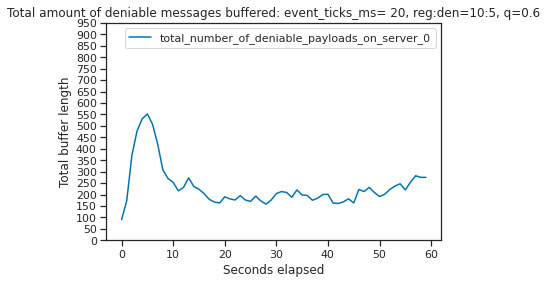}
         \caption{Deniable buffer length over time.}
         \label{fig:graph-buffer-a6}
     \end{subfigure}
\caption{Server statistics collected with setting A6.}\label{fig:graph-server-statistics-a6}
\end{figure}

\begin{figure}[htb]
     \centering
     \begin{subfigure}[b]{0.3\textwidth}
         \centering
         \includegraphics[width=\textwidth]{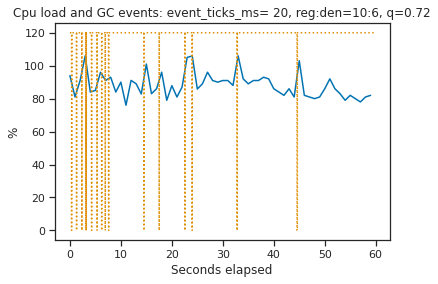}
         \caption{CPU load over time.}
         \label{fig:graph-cpu-a7}
     \end{subfigure}
     \hfill
     \begin{subfigure}[b]{0.31\textwidth}
         \centering
         \includegraphics[width=\textwidth]{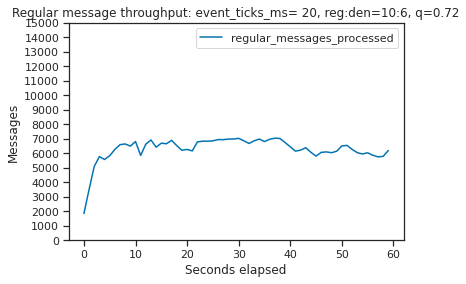}
         \caption{Regular messages processed over time.}
         \label{fig:graph-throughput-a7}
     \end{subfigure}
     \begin{subfigure}[b]{0.37\textwidth}
         \centering
         \includegraphics[width=\textwidth]{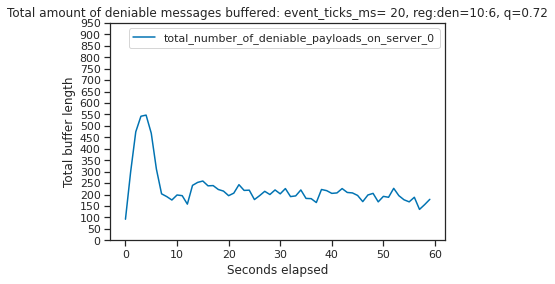}
         \caption{Deniable buffer length over time.}
         \label{fig:graph-buffer-a7}
     \end{subfigure}
\caption{Server statistics collected with setting A7.}\label{fig:graph-server-statistics-a7}
\end{figure}

\begin{figure}[htb]
     \centering
     \begin{subfigure}[b]{0.3\textwidth}
         \centering
         \includegraphics[width=\textwidth]{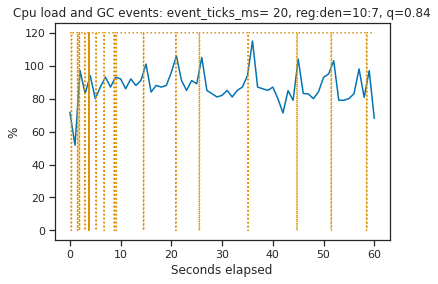}
         \caption{CPU load over time.}
         \label{fig:graph-cpu-a8}
     \end{subfigure}
     \hfill
     \begin{subfigure}[b]{0.31\textwidth}
         \centering
         \includegraphics[width=\textwidth]{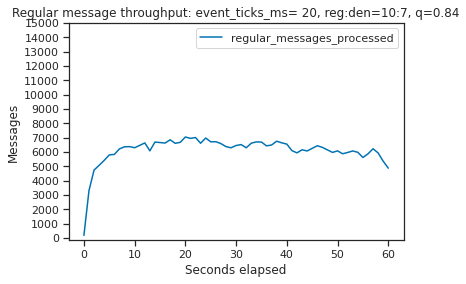}
         \caption{Regular messages processed over time.}
         \label{fig:graph-throughput-a8}
     \end{subfigure}
     \begin{subfigure}[b]{0.37\textwidth}
         \centering
         \includegraphics[width=\textwidth]{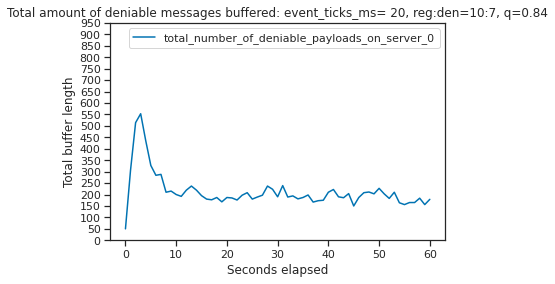}
         \caption{Deniable buffer length over time.}
         \label{fig:graph-buffer-a8}
     \end{subfigure}
\caption{Server statistics collected with setting A8.}\label{fig:graph-server-statistics-a8}
\end{figure}

\begin{figure}[htb]
     \centering
     \begin{subfigure}[b]{0.3\textwidth}
         \centering
         \includegraphics[width=\textwidth]{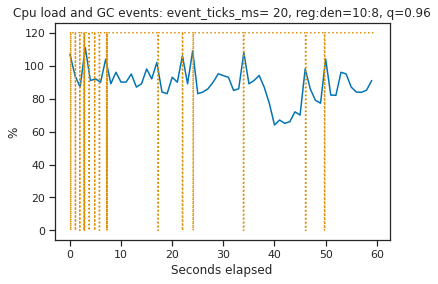}
         \caption{CPU load over time.}
         \label{fig:graph-cpu-a9}
     \end{subfigure}
     \hfill
     \begin{subfigure}[b]{0.31\textwidth}
         \centering
         \includegraphics[width=\textwidth]{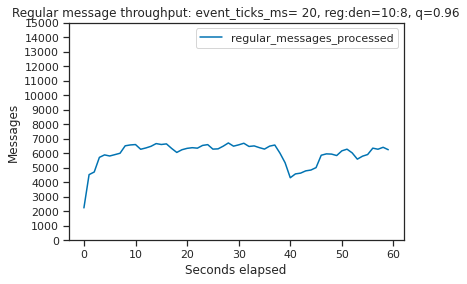}
         \caption{Regular messages processed over time.}
         \label{fig:graph-throughput-a9}
     \end{subfigure}
     \begin{subfigure}[b]{0.37\textwidth}
         \centering
         \includegraphics[width=\textwidth]{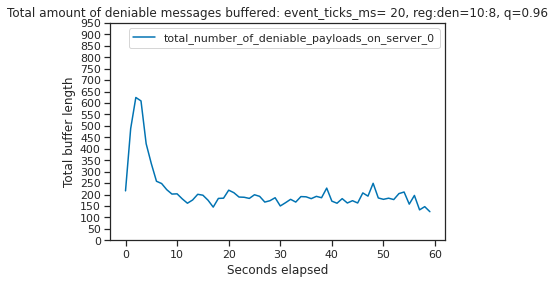}
         \caption{Deniable buffer length over time.}
         \label{fig:graph-buffer-a9}
     \end{subfigure}
\caption{Server statistics collected with setting A9.}\label{fig:graph-server-statistics-a9}
\end{figure}

\begin{figure}[htb]
     \centering
     \begin{subfigure}[b]{0.3\textwidth}
         \centering
         \includegraphics[width=\textwidth]{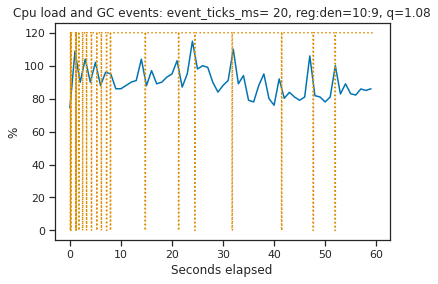}
         \caption{CPU load over time.}
         \label{fig:graph-cpu-a10}
     \end{subfigure}
     \hfill
     \begin{subfigure}[b]{0.31\textwidth}
         \centering
         \includegraphics[width=\textwidth]{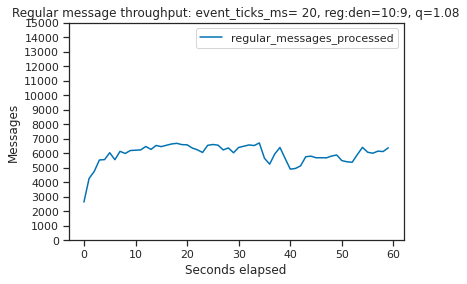}
         \caption{Regular messages processed over time.}
         \label{fig:graph-throughput-a10}
     \end{subfigure}
     \begin{subfigure}[b]{0.37\textwidth}
         \centering
         \includegraphics[width=\textwidth]{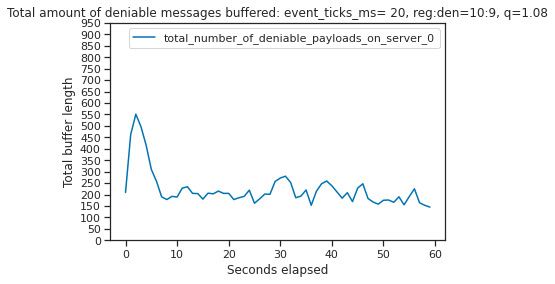}
         \caption{Deniable buffer length over time.}
         \label{fig:graph-buffer-a10}
     \end{subfigure}
\caption{Server statistics collected with setting A10.}\label{fig:graph-server-statistics-a10}
\end{figure}

\begin{figure}[htb]
     \centering
     \begin{subfigure}[b]{0.3\textwidth}
         \centering
         \includegraphics[width=\textwidth]{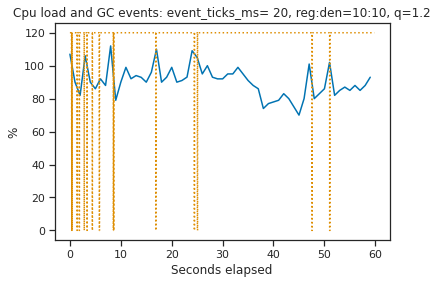}
         \caption{CPU load over time.}
         \label{fig:graph-cpu-a11}
     \end{subfigure}
     \hfill
     \begin{subfigure}[b]{0.31\textwidth}
         \centering
         \includegraphics[width=\textwidth]{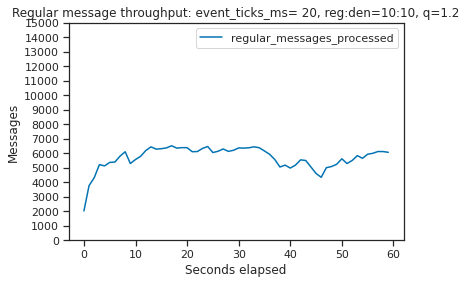}
         \caption{Regular messages processed over time.}
         \label{fig:graph-throughput-a11}
     \end{subfigure}
     \begin{subfigure}[b]{0.37\textwidth}
         \centering
         \includegraphics[width=\textwidth]{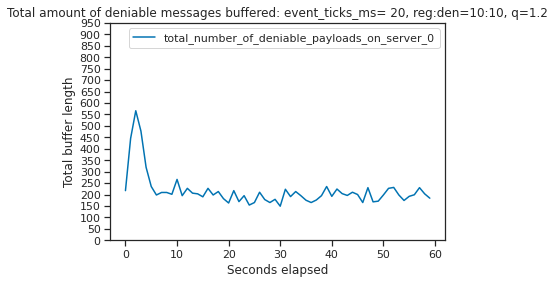}
         \caption{Deniable buffer length over time.}
         \label{fig:graph-buffer-a11}
     \end{subfigure}
\caption{Server statistics collected with setting A11.}\label{fig:graph-server-statistics-a11}
\end{figure}

\begin{figure}[htb]
     \centering
     \begin{subfigure}[b]{0.45\textwidth}
         \centering
         \includegraphics[width=0.8\textwidth]{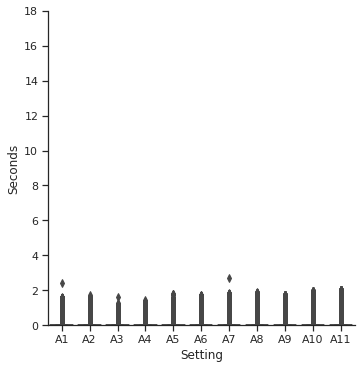}
         \caption{Regular message latency over time.}
         \label{fig:graph-regular-latency}
     \end{subfigure}
     \hfill
     \begin{subfigure}[b]{0.45\textwidth}
         \centering
         \includegraphics[width=0.8\textwidth]{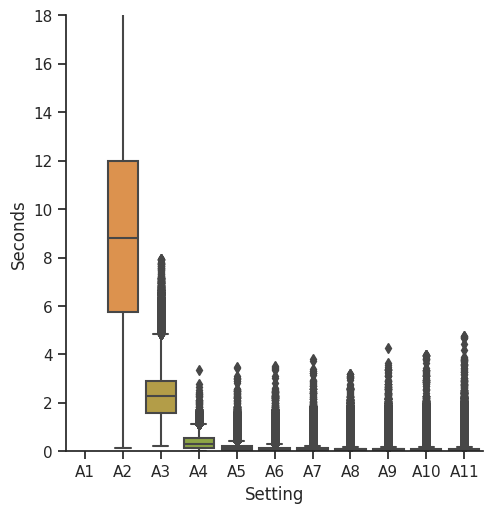}
         \caption{Deniable messages latency over time.}
         \label{fig:graph-deniable-latency}
     \end{subfigure}
\caption{Client-to-client message latency measure from $\padding=0$ to $\padding=1.2$ with step 0.12.}\label{fig:graph-latency-same-scale}
\end{figure}

\FloatBarrier

\section{Full systematization table}\label{app:rw-full-table}
Detailed table of related work and this work is presented in \Cref{tab:app-rw-comparison-full}.

\begin{table*}[tb]
    \centering
    \begin{tabularx}{\linewidth}{llllXX}
        \toprule
    \textbf{Protocol} & \textbf{Tunneling} & \textbf{Censorship resilience} &\textbf{Provable guarantees} & \textbf{Trust} & \textbf{Threat model}  \\
    \midrule

        DC-nets~\cite{chaum_dining_1988} & \xmark & Weak & Transport layer & Anytrust & GA \\
        Dissent~\cite{corrigan-gibbs_dissent_2010} & \xmark & Weak & Transport layer & Anytrust & GA \\
        Anonycaster~\cite{head_anonycaster_2012} & \xmark & Weak & Transport layer & Anytrust & GA \\
        Riffle~\cite{kwon_riffle_2016} & \xmark & Weak & Transport layer & Anytrust & GA \\ 
        Atom~\cite{kwon_atom_2017} & \xmark & Weak & Transport layer & Anytrust & GA \\ 
        Talek~\cite{cheng_talek_2020} & \xmark & Weak & Transport layer & Anytrust & GA \\

        Herd~\cite{le_blond_herd_2015} & \xmark & Weak & Transport layer & Chosen set & GP/LA \\ 

        Pynchon~\cite{sassaman_pynchon_2005}  & \xmark & Weak & Transport layer & Fraction & GA\\ 
        XRD~\cite{kwon_xrd_2020} & \xmark & Weak & Transport layer & Fraction & GA\\
        Express~\cite{eskandarian_express_2021} & \xmark& Weak & Transport layer & Fraction & GA\\ 

        $P^3$~\cite{kissner_private_2004} & \xmark & Weak & Transport layer & Honest-but-curious/Malicious & GA \\ 
        Loopix~\cite{piotrowska_loopix_2017} & \xmark & Weak & Transport layer & Honest-but-curious & GA \\ 

        Bitmessage~\cite{warren_bitmessage_2012}  & \xmark & Weak & Transport layer & Zero-trust & GA \\ 
        Pung~\cite{angel_unobservable_2016} & \xmark & Weak & Transport layer & Zero-trust & GA\\ 
        Riposte~\cite{corrigan-gibbs_riposte_2015} & \xmark & Weak & Transport layer &  Zero-trust*** & GA\\ 
        

        Verdict~\cite{corrigan-gibbs_proactively_2013} & \xmark & Weak & Transport layer* & Anytrust & GA \\
        Stadium~\cite{tyagi_stadium_2017} & \xmark & Weak &  Transport layer* & Anytrust & GA\\ 
        
        Vuvuzela~\cite{van_den_hooff_vuvuzela_2015} & \xmark & Weak &  Transport layer* & Fraction & GA \\
        Alpenhorn~\cite{lazar_alpenhorn_2016} & \xmark & Weak &  Transport layer* & Fraction & GA \\ 
        Karaoke~\cite{lazar_karaoke_2018} & \xmark & Weak &  Transport layer* & Fraction & GP \\ 
        Yodel~\cite{lazar_yodel_2019} & \xmark & Weak & Transport layer** & Fraction & GA\\ 
        
        Groove~\cite{barman_groove_2022} & \xmark & Weak & Transport layer* & Zero-trust & GA \\ 


        Mixminion~\cite{danezis_mixminion_2003} & \xmark & Weak & \noguarantee & Chosen path & GA\\ 
        HORNET~\cite{chen_hornet_2015} & \xmark & Weak & \noguarantee & Fraction & LA \\ 
        Tor~\cite{dingledine_tor_2004} & \xmark &  Weak & \noguarantee & One**** & GP \\


        \midrule
        \ourprotocol & \cmark & Strong & Application layer & Centralized & GA \\
        \midrule


        CoverDrop~\cite{ahmed-rengers_coverdrop_2022} & \cmark & Strong & \noguarantee & Centralized & GP \\ 
        
        Cirripede~\cite{houmansadr_cirripede_2011} & \cmark & Strong & \noguarantee & Proxies & LA \\
        Telex~\cite{wustrow_telex_2011} & \cmark & Strong & \noguarantee & Proxies & LA\\ 
        CensorSpoofer~\cite{wang_censorspoofer_2012} & \cmark & Strong & \noguarantee & Proxies & LA \\ 
        FreeWave~\cite{houmansadr_i_2013}  & \cmark & Strong & \noguarantee & Proxies & LA\\ 

        SkypeMorph~\cite{mohajeri_moghaddam_skypemorph_2012} & \cmark & Strong & \noguarantee & Proxy & LA \\
        IMProxy~\cite{bahramali_practical_2020} & \cmark & Strong & \noguarantee & Proxy & LA\\
        Protozoa~\cite{barradas_poking_2020} & \cmark & Strong & \noguarantee & Proxy & LA \\
        Camoufler\cite{sharma_camoufler_2021} & \cmark & Strong & \noguarantee & Proxy & LA  \\
        
        Balboa~\cite{rosen_balboa_2021} & \cmark & Strong & \noguarantee & Zero-trust & GA \\

    \bottomrule
    \end{tabularx}
    \caption{Comparison of related work. Footnotes: *via differential privacy, **with failure probability $10^{-8}$ per round, ***for privacy, all servers need to be trusted for availability, ****in chosen set. G=Global, L=Local, A=Active, P=Passive.}
    \label{tab:app-rw-comparison-full}
\end{table*}

\else

\begin{abstract}
Transport layer data leaks metadata unintentionally -- such as who communicates with whom. 
While tools for strong transport layer privacy exist, they have adoption obstacles, including performance overheads incompatible with mobile devices.
We posit that by changing the objective of metadata privacy for \emph{all traffic}, we can open up a new design space for pragmatic approaches to transport layer privacy.
As a first step in this direction, we propose using techniques from information flow control and present a principled approach to constructing formal models of systems with metadata privacy for \emph{some}, deniable, traffic.
We prove that deniable traffic achieves metadata privacy against strong \attackers -- this constitutes the first bridging of information flow control and anonymous communication to our knowledge.
Additionally, we show that existing state-of-the-art protocols can be extended to support metadata privacy, by designing a novel protocol for \emph{deniable instant messaging} (\ourprotocol), which is a variant of the Signal protocol.
To show the efficacy of our approach, we implement and evaluate a proof-of-concept instant messaging system running \ourprotocol on top of unmodified Signal. 
We empirically show that the \denimonsignal can maintain low-latency for unmodified Signal traffic without breaking existing features, while at the same time supporting deniable Signal traffic.
\end{abstract}

\pagestyle{plain}
\thispagestyle{empty}
\section{Introduction}\label{sec:introduction}
Modern instant messaging (IM) services strive for strong end-to-end security. Services such as Signal, WhatsApp~\cite{whatsapp_whatsapp_2020}, Wire~\cite{wire_swiss_gmbh_wire_2021}, and Facebook Messenger~\cite{facebook_newsroom_messenger_2016}, all use the Signal protocol that is formally secure~\cite{cohn2020formal} and achieves ambitious security goals, such as post-compromise security, backward secrecy, confidentiality, and integrity.
Still, these IM services lack strong metadata privacy, making them vulnerable to traffic analysis attacks. 
This is a serious deficiency, because traffic analysis remains an effective mechanism~\cite{fu_service_2016,taylor_robust_2018} for surveillance and censorship used by governments, organizations, and internet service providers in over 100 countries~\cite{raman_measuring_2020}. 
For example, China's ``great firewall'' actively probes and censors privacy tools~\cite{ensafi_examining_2015}. Although collecting metadata may seem non-intrusive, metadata is used to make critical decisions -- ``we kill people based on metadata'', as former US government official general Hayden~\cite{johns_hopkins_university_johns_2014} put it. 
``Harvest today, analyze tomorrow'' is a viable adversarial strategy for many such actors.

While the general problem of metadata privacy has been extensively studied, there are both social and technical barriers that prevent adoption of existing privacy tools to IM services.
On a social level, people are either unaware of privacy tools~\cite{ruogu_data_goes_everywhere_2015} or have diverse misconceptions about them~\cite{story_awareness_2021}. 
Adding to the existing problem, people also find these tools too complicated to use, or lack the knowledge of how to use them~\cite{gerber_why_2019}. For example, Norcie et al.~\cite{norcie_why_2014} investigated the Tor Browser Bundle and found that users experienced usability issues such as the browser's launch time and difficulties downloading and installing it. 
Beyond the challenges of usability, there are risks of being scrutinized for having a particular app installed~\cite{samuel_china_2019,wagstaff_failing_2013}. 

On a technical level, available tools are far from perfect. 
The Tor project~\cite{tor_project_tor_nodate} although relatively popular with 2M active users~\cite{the_tor_project_users_2021}, is vulnerable to de-anonymization~\cite{karunanayake_-anonymisation_2021}, denial of service (DoS)~\cite{jansen_point_2019}, and traffic analysis~\cite{nasr_deepcorr_2018}. 
Because Tor can be automatically fingerprinted~\cite{fu_service_2016,taylor_robust_2018}, it is also easy to block (ironically, the authors of this paper themselves were blocked from accessing the Tor project's website on their organization network). 
Metadata private focused IM tools that run on Tor~\cite{zbay_how_nodate-1,briar_how_nodate,cwtch_overview_nodate,ricochet_refresh_nodate} suffer from the same issues. 
Other tools that hide traffic by imitating well-known apps do not produce credible traffic~\cite{houmansadr_parrot_2013}. 

The strongest guarantees for metadata privacy are provided by dedicated protocols. 
In particular, round-based, DC-nets like, protocols~\cite{wolinsky_dissent_2012,corrigan-gibbs_dissent_2010,van_den_hooff_vuvuzela_2015,lazar_karaoke_2018}, where predetermined rounds make traffic patterns indistinguishable, are able to resist traffic analysis. 
However, round-based protocols are both resource exhaustive and inflexible. 
The rounds themselves require constant overhead, which results in poor performance~\cite{gilad_metadata-private_2019}, making them especially infeasible for resource constrained devices such as phones or wearables. 
Moreover, a major obstacle with round-based protocols is that they depend on fixed sets of individuals participating. That is, participants cannot join or leave without changing the privacy guarantees. Finally, round-based protocols are also easy to fingerprint and block.

Existing approaches to metadata privacy all have in common that they focus on the strong objective of \emph{metadata privacy for all users all the time}. 
However, such a strong objective significantly delimits the design space of possible solutions. 
We propose a different, pragmatic objective: \textbf{rather than offering privacy to all users all the time, let us offer privacy to all users some of the time}. 
This shift in objective expands the design space for metadata privacy to new solutions.

Our new approach is to incorporate metadata privacy into an existing store-and-forward IM protocol.
To that extent, we present \ourprotocollong -- an IM protocol that provides both message confidentiality and metadata privacy.
\ourprotocol\ distinguishes two kinds of messages: (i) \emph{regular} messages that do not require metadata privacy, and (ii) \emph{deniable} messages that do require it.
Regular and deniable communication is combined in one system, and 
users decide which messages to send privately.
To withstand traffic analysis, deniable messages are not communicated immediately, instead they are piggybacked on top of the regular messages, which in turn requires that all messages are extended by a small known amount of bytes. The store-and-forward server breaks the link between the sender and receiver of a private message by buffering the message until there is an opportunity to piggyback it on some other regular message to the receiver.
It is vital that the messages are extended even if the communicating parties have nothing to say, in which case a dummy payload is sent instead. To minimize overhead, the size of the payload must be small in proportion to the overall communication.

The importance of incorporating metadata privacy into an existing IM protocol aligns with earlier observations in the literature. 
As EFF put it, ``An app with great security features is worthless if none of your friends and contacts use it''~\cite{electronic_frontier_foundation_communicating_2020}. 
In their paper ``\emph{Practical Traffic Analysis Attacks on Secure Messaging Applications'}', Bahramali et al.~\cite{bahramali_practical_2020} recommend that metadata privacy for IM should be adopted by IM services to be effective. 
An encouraging development in this direction already is WhatsApp's use of the Noise Protocol Framework (NPF)~\cite{perrin_noise_2018} to protect certain metadata~\cite{whatsapp_whatsapp_2020}.  
Finally, Zuckerman's~\cite{zuckerman_cute_2015} \emph{cute cat theory of censorship} posits that platforms that combine entertainment with political activism are more resilient to censorship than dedicated political platforms. 

In our case, we  pair \ourprotocol\ with the (unmodified) Signal protocol. 
We call the resulting system \emph{\denimonsignal}. We chose Signal  because it provides the state-of-the art security guarantees in instant messaging, including: \emph{forward} secrecy, \emph{backward} secrecy (post-compromise security), data \emph{confidentiality} and \emph{integrity} (see \cite{cohn2020formal} for detail).

\ourprotocol's piggybacking of deniable messages is a form of tunneling (e.g., \cite{houmansadr_i_2013,barradas_poking_2020,sharma_camoufler_2021}). Yet
tunneling alone is insufficient. The reason is that in settings where adversaries are legitimate users in the system, information may be leaked inadvertently through parts of the protocol state that are shared between all users. 
For example, in Signal, adversaries can gain information about other users through the state of the key distribution center because the protocol allow users to run out of keys -- this allows adversaries to count the number of keys a user has and in extension allows the adversary to deduce how many conversations a user is part of.

To ensure that \ourprotocol\ guarantees metadata privacy for deniable messages, including unknown attacks, we use techniques from secure information flow. 
Our insight is to model user deniable behavior as \emph{user strategies}~\cite{infoflow:strategies} -- a technical device that is traditionally used for specifying semantic security of interactive and nondeterministic programs. 
In \ourprotocol, a user strategy is a function that given a history of the user's communication determines their next deniable action, e.g., send a deniable message, request key material from the server to initiate new deniable communication, or block a user from receiving deniable messages.
We recast the notion of metadata privacy as strategy-based noninterference: \emph{user strategies must not leak through the protocol}.
The significance of this insight is that because noninterference is an end-to-end characterization, proving noninterference requires that there is no way in which the sensitive information may leak anywhere in the protocol, not just on the transport layer. In essence, this guides the features and non-features of \ourprotocol. For example, \ourprotocol\ restricts the notification of user blocking, because notifying a user that they have been blocked leaks information about the blocking user's deniable behavior.

The contributions of this paper are as follows:
\begin{itemize}[noitemsep,nolistsep]
    \item It presents a deniable variant of the Signal protocol (\Cref{sec:signal}) called \ourprotocol (\Cref{sec:denim-details}), that supports both the original strong cryptographic guarantees of Signal, and metadata privacy.
    \item It presents a system design that layers deniable Signal messages on top of the unmodified Signal protocol, which we call \denimonsignal (\Cref{sec:denim-on-signal}).
    \item It presents a formal privacy analysis (\Cref{sec:formal-model}) that constitutes a principled approach of using information flow techniques to guarantee privacy by proving noninterference. 
    \item It presents a proof-of-concept implementation (\Cref{sec:implementation}) of an instant messaging system with \ourprotocol.
    \item It presents an empirical evaluation (\Cref{sec:experiment-setup,sec:experiment-results}) of the performance of \ourprotocol.
\end{itemize}

\section{Background}\label{sec:background}
This section provides an overview of instant messaging (IM), and the main machinery in the Signal protocol which we design a deniable version of in \Cref{sec:denim-details}.

\subsection{Instant messaging}\label{sec:instant-messaging}
In 2019, instant messaging (IM) services had seven billion registered accounts worldwide~\cite{the_radicati_group_inc_instant_2019}. The most popular IM services include WhatsApp (2B users), Facebook messenger (1.3B users), iMessage (estimated to 1B users), Telegram (550M users), and Snapchat (538M users)~\cite{statista_most_2021,kastrenakes_apple_2021}. While IM appears deceptively simple, the sheer amount of users and traffic (69M messages/min in 2021~\cite{jenik_heres_2021}) present several engineering challenges. Keeping up with the demands, requires deploying and maintaining robust systems. As an example, WhatsApp's architecture handles over one million connections per server~\cite{ixsystems_inc_rick_2014}.

\ifverbose
\subsection{Architecture}\label{sec:IM-architecure}
All major IM services, including WhatsApp, Facebook messenger, Telegram and Snapchat, use centralized servers to forward messages between clients~\cite{bahramali_practical_2020}. To exemplify, WhatsApp use centralized servers to forward chat messages, and use dedicated pull servers only for larger attachments such as pictures and videos~\cite{ixsystems_inc_rick_2014}. Similarly, iMessage also explicitly use centralized servers via their Apple Push Notification service (APNs), to forward messages between clients. Still, IM services rarely provide full specifications of their network topology.

\subsection{Encryption schemes}\label{sec:IM-encryption}
Today, many IM apps come with end-to-end encryption. Telegram uses their own protocol, MTProto~\cite{telegram_mtproto_nodate}, iMessage uses RSA encryption with optimal asymmetric encryption padding (OAEP)~\cite{apple_inc_how_2021}, and Snapchat uses an unnamed encryption scheme for some of its content~\cite{salim_finally_2019}. WhatsApp~\cite{whatsapp_whatsapp_2020} and Facebook Messenger~\cite{facebook_newsroom_messenger_2016} on the other hand both use the Signal protocol~\cite{marlinspike_advanced_2013,marlinspike_x3dh_2016}. 
Of these protocols, Signal is the most popular, as it is also used by Wire~\cite{wire_swiss_gmbh_wire_2021}, ChatSecure, Conversations, Pond, the Signal app, and Silent Circle~\cite{cohn-gordon_formal_2020}. The Signal protocol has also been shown to be formally secure~\cite{cohn-gordon_formal_2020}.
Signal is based on Off-the-Record Messaging (OTR)~\cite{borisov_off--record_2004} and Silent Circle Instant Messaging Protocol (SCIMP)~\cite{moscaritolo_silent_2012}.

\subsection{Metadata privacy}\label{sec:metadata-privacy-IM}
WhatsApp has made a conscious effort to protect users' metadata. Specifically, WhatsApp uses the Noise Protocol Framework (NPF)~\cite{perrin_noise_2018} to encrypt metadata~\cite{whatsapp_whatsapp_2020}. NPF aims to prevent identity leakage by encrypting the public key of a sender. As such, in WhatsApp, NPF hides the public key of the sender from the server. Still, NPF is not resistant to traffic analysis, and does not hide other metadata such as IP addresses.
\else
All major IM services, including WhatsApp, Facebook messenger, Telegram and Snapchat, use centralized servers to forward messages~\cite{bahramali_practical_2020}. 
Many IM apps also come with end-to-end encryption, in addition to server-client encryption (through TLS). 
Telegram uses their own protocol, MTProto~\cite{telegram_mtproto_nodate}, iMessage uses \rsaencryption~\cite{apple_inc_how_2021}, and Snapchat uses an unnamed encryption scheme for some of its content~\cite{salim_finally_2019}. 
The most popular protocol is Signal~\cite{marlinspike_advanced_2013,marlinspike_x3dh_2016}, which also has the strongest security guarantees of the mentioned protocols, and is used by WhatsApp~\cite{whatsapp_whatsapp_2020}, Facebook Messenger~\cite{facebook_newsroom_messenger_2016}, Wire~\cite{wire_swiss_gmbh_wire_2021}, ChatSecure, Conversations, Pond, the Signal app, and Silent Circle~\cite{cohn2020formal}. 
The Signal protocol is formally secure~\cite{cohn2020formal}, and is based on Off-the-Record Messaging (OTR)~\cite{borisov_off--record_2004} and Silent Circle Instant Messaging Protocol (SCIMP)~\cite{moscaritolo_silent_2012}. 
Despite strong cryptographic guarantees, none of the centralized IM services support transport layer privacy for IM.
\fi

\subsection{The Signal protocol}\label{sec:signal}
At a high level the Signal protocol realizes an end-to-end secure communication channel between two parties that exchange instant messages in a possibly asynchronous way (i.e., they may not be online at the same time). 
Signal distinguished itself among the landscape of messaging protocols in that it achieves ambitious security goals including: \emph{forward} secrecy, \emph{backward} secrecy (post-compromise security), data \emph{confidentiality} and \emph{integrity} (see \cite{cohn2020formal} for detail).  This is obtained by managing several different cryptographic keys (\Cref{tab:signal-keys}), relying on a semi-trusted centralized server (to store and forward messages, and implement a key distribution center), 
and 
cleverly combining three cryptographic primitives: a key derivation function (KDF), a non-interactive key-exchange protocol (namely DH for Diffie-Hellman) for initiating new sessions, and an authenticated encryption scheme with associated data (AEAD). 

\subsubsection{Keys used in Signal}\label{sec:signal-keys}
In Signal, each user $U$ holds a set of keys that identify the user, and are used to initiate new sessions (chats) and to AEAD-encrypt messages.
\Cref{tab:signal-keys} provides a categorization of the cryptographic key material of Signal that is relevant to this work. Keys employed only to set up new sessions are highlighted with the symbol $^\bigstar$.

\begin{table}[htb]
    \centering
    \scalebox{1}{
    \begin{tabularx}{1.\linewidth}{llX}
    \toprule
        \textbf{Name} & \textbf{Key(s)} & \textbf{Usage} \\\midrule
        Identity key-pair$^\bigstar$ & \{${\longtermpublickey}_U, \mathit{\longtermprivatekey}_U$\} & Long-term\\
        Mid-term key-pair$^\bigstar$ & \{$\mathit{\midtermpublickey}_U, \mathit{\midtermprivatekey}_U$\} & Mid-term\\
        Ephemeral key-pair$^{(\bigstar)}$ & \{$\mathit{\ephemeralpublickey}_U, \mathit{\ephemeralprivatekey}_U$\} & One-time\\
        Master secret & \ms & One-time\\
        Message key & $\mathit{\messagekey_{x,y}}$ & One-time\\\bottomrule
    \end{tabularx}
    }
    \caption{List of Signal's keys that are relevant to this work. Ephemeral keys are used in various parts of the Signal protocol, when employed in session initialization they are commonly called one-time keys.}
    \label{tab:signal-keys}
\end{table}

\subsubsection{Overview of the Signal Protocol}\label{sec:signal-overview}
What follows recalls the essential facts needed to understand this work (a full formalization of Signal is available in \cite{perrin_double_2016,cohn2020formal}). 
The Signal protocol is made of three main steps:

\ourparagraph{User registration}
Run once in the lifetime of a user in the system. This step entails storing a user's public key material in the Signal server, namely $\longtermpublickey_U, \midtermpublickey_U$, and a set of (one-time) ephemeral public keys $\{\ephemeralpublickey_U^{(1)},\ldots,\ephemeralpublickey_U^{(n)}\}$.

\ourparagraph{New-session initialization}
Run once per new session initiated by the user. This step is used to start a new chat. The requesting user $A$ interacts with the server to obtain the handle of $B$, another user, consisting of $\longtermpublickey_B, \midtermpublickey_B$ and a single one-time public key $\ephemeralpublickey_B^{(i)}$. $A$ uses $B$'s keys together with their identity secret key, long term secret key and an ephemeral secret key to run a non-interactive key-exchange and generate a master secret key $\ms_{AB}$, that is computable only by $A$ and $B$.

\ourparagraph{The {double ratchet} mechanism for messaging}
Run every time the user receives or sends a new message. In Signal every message is AEAD-encrypted under a different message key $\messagekey_{x,y}$. We index the message keys by two non-negative integers $x,y$ that operate as coordinates. The value $x$ identifies the current sender, $y$ the number of messages sent by the current sender since the last change of speaker. Thus even values of $x$ correspond to events where the current speaker is the initiator of the chat, while $y$ denotes how many messages the sender of level $x$ has sent so far. In order to securely derive new keys from previous ones, the double ratchet mechanism ingeniously combines two KDFs.

\section{System design}\label{sec:design}
This section presents the scope and goals of our deniable messaging system, \emph{\denimonsignal}.
We start by defining the threat model, which will dictate the necessary design goals and trust assumptions.

\subsection{Threat model}\label{sec:adv-model}
We consider a global active \attacker\ who participates in the deniable protocol. 
The \attacker\ can: 
\begin{itemize}[noitemsep,nolistsep]
    \item Observe the entire network, including messages to and from the server, and to and from the users.
    \item Insert or modify traffic.
    \item Participate in the protocol. This gives to the \attacker\ access to the parts of the protocol state that are accessible to all protocol participants, including requesting other users' keys from the key distribution center (KDC), and sending messages.
\end{itemize}
The \attacker\ cannot compromise the internal state of honest parties, including servers.

Under this threat model, the \attacker\ could for example be an internet service provider, or a nation-state.
Given these capabilities, the goal of the \attacker\ is:
\begin{itemize}[noitemsep,nolistsep]
    \item To learn or to alter the payload of deniable traffic between honest parties. 
    \item To learn whether a given network message contains deniable payload or not. 
    \item To learn whether two parties have an ongoing exchange of deniable traffic or not.
\end{itemize}

Note that our system makes traffic 'deniable' on the transport layer, which is different from e.g., deniable encryption~\cite{canetti_deniable_1997} where the goal is to give deniability for the message content (plaintext) rather than hiding fact of communication.

\subsection{Design goals}\label{sec:goals}
The high-level goal of our system is to be resilient against \attackers\ with the goals in \Cref{sec:adv-model}.
Additionally, tunneling deniable traffic inside instant messaging systems requires making decisions regarding performance trade-offs between the deniable traffic and the regular traffic. 
We derive the design goals for security and privacy (\Cref{sec:security-and-privacy-goals}) based on the threat model and instant messaging use case, and design goals for performance (\Cref{sec:performance-goals}) from the use case.

\subsubsection{Security and privacy goals}\label{sec:security-and-privacy-goals}
\hfill\\
\ourparagraph{Confidentiality of users' deniable behavior}
A consequence of our threat model is that an \attacker\ could try to infer users' deniable behavior both by observing the network, or by observing shared protocol states. Adequate protection measures therefore depend on data the users generate by interacting with the deniable protocol not leaking into channels the \attacker\ can observe (the network and the shared state). 
That is, a successful implementation depends on proving noninterference between the deniable protocol and the protocol it piggybacks on -- noninterference ensures all of the users' input to the deniable protocol is kept confidential, not just that the network traffic is protected.

\ourparagraph{Privacy guarantees independent of the number of online users} 
To achieve strength-in-numbers, we aim for the design
where the privacy guarantees do not depend on the dynamic behavior of the system, i.e., users may join or leave the system without significantly affecting the privacy of others.
This means that the system should tunnel the deniable traffic using an observable protocol that does not achieve transport layer privacy on its own.

\ourparagraph{Strong security guarantees for deniable messages} Message content should be protected using state-of-the-art techniques, which for IM can be achieved via the Signal protocol. Signal is more than just a mere key exchange protocol; it is designed to deliver not only confidentiality and integrity, but also more advanced security features such as key healing. We aim to maintain the same security benefits provided by Signal by carefully building our deniable messaging machinery around the Signal protocol in a way that does not impact Signal's security functionalities.

\subsubsection{Performance goals}\label{sec:performance-goals}
\hfill\\
\ourparagraph{Parameterizable bandwidth overhead for deniable traffic} 
To control the privacy-performance trade-offs in the system, the deniable payload overhead should be a global tuneable parameter that is set on a case-by-case basis to match a user population's demand for deniable traffic. There should be no limitation on the length of the regular traffic.

\ourparagraph{Prioritize low latency for regular traffic} We prioritize the performance of regular traffic -- it is important that users continue to use the regular IM system -- above the performance of the deniable traffic. This creates an asymmetry in the latency of regular and deniable communication. When using the protocol for regular Signal, traffic is forwarded immediately resulting in low latency overhead. For \ourprotocol, the latency depends on when traffic can be safely piggybacked. While a system with different privacy guarantees for different messages like this has not yet been studied from the usability perspective, we assume that a higher latency overhead for deniable communication is tolerable as the privacy guarantees are stronger.

\subsection{Trust assumptions}\label{sec:assumptions}
The previously stated design goals, combined with the threat model, leads to the following trust assumptions:
\begin{itemize}[noitemsep,nolistsep]
    \item The \attacker\ cannot access the internal state of honest parties.
    \item Users trust receivers of their deniable traffic, i.e. users are by design not able to deny having sent traffic to their intended receiver.
    \item Users' deniable behavior does not influence their regular behavior, e.g., a user does not send more regular traffic than they normally would to piggyback their deniable traffic.
    \item The forwarding servers are trusted.
    \item The KDC is trusted, and can generate ephemeral keys on behalf of a user in case the user's deniable ephemeral keys have been depleted.
    \item Users do not issue deniable key requests for adversaries' keys, and do not respond to deniable Signal sessions initiated by adversaries.
\end{itemize}

Note that our trust assumptions to a large extent are inherited from the use case, IM, and from the Signal protocol.
For example, centralized, trusted servers is the natural setting for IM.
Moreover, Signal assumes that a user is able to verify that the receiver of messages is a trusted party using an out of bounds channel -- in their deployment they support this by providing a QR code that both parties are supposed to verify in person.

Our formal model (\Cref{sec:formal-model}) incorporates the trust assumptions at a technical level.

\section{DenIM on Signal}\label{sec:denim-on-signal}

This section presents \emph{\denimonsignal}, an instant messaging system that supports two different protocols: regular Signal, and our deniable variant of Signal, \ourprotocollong.
\ourprotocol is a centralized IM protocol with both the cryptographic guarantees of the Signal protocol, and transport layer privacy for messages. 
In \denimonsignal the deniable protocol, \ourprotocol, piggybacks on an unmodified version of Signal.

At a high level, \denimonsignal provides users with two communication abstractions: sending `regular' Signal traffic that is not resilient to traffic analysis, and sending `deniable' Signal messages that come with transport layer privacy. 
To prevent an \attacker\ from trivially inferring which users are communicating, \denimonsignal\ uses a simple centralized architecture where traffic is routed through a trusted server. 
The server forwards the regular Signal traffic immediately, and stores the \ourprotocol\ traffic until there is regular traffic for the intended recipient to piggyback on. 
To prevent an \attacker\ from tracing traffic by fingerprinting it as it is forwarded by the server, the traffic between clients and server is sent over TLS.

We model and prove the security of our implementation in \Cref{sec:formal-model}, and empirically evaluate how bandwidth overhead affects system performance both for deniable and regular traffic in \Cref{sec:evaluation}.

\subsection{Protocol details}\label{sec:denim-details}
In this section we elaborate on the technical details of \ourprotocol. 
We explain how the deniable part of a network message is created (\Cref{sec:padding}), how and where deniable parts get buffered (\Cref{sec:buffers}), and which content can be carried in the deniable part i.e., what deniable actions are supported by \ourprotocol\ (\Cref{sec:functionality}). 
\ourprotocol is a variant of Signal -- we make a minor change to the Signal protocol (\Cref{sec:signal-changes}) to ensure that \ourprotocol is a deniable variant of Signal (the standard Signal protocol would otherwise leak information about a user's deniable sessions), but otherwise encapsulates Signal. 
We stress that this modification does not impact the cryptographic security.  

\ourparagraph{Communication flow by example}
\Cref{fig:our-protocol} presents an example of communication flow in \denimonsignal. The purple user (upper left) has queued a deniable message (D) waiting to be sent to the orange user (lower right). 
As purple sends a regular message (R) to the green user (upper right), part of their deniable message for the orange user is added to the deniable padding. 
The server immediately forwards the regular message to green, and as there are no deniable messages queued for green, the message is padded with dummy padding. 
Next, the blue user (lower left) sends a regular message to orange. 
The server forwards the regular message to orange, and adds purple's deniable message that has been waiting on the server to the padding of the message for orange.

\begin{figure}[tb]
    \centering
    \includegraphics[width=0.96\columnwidth]{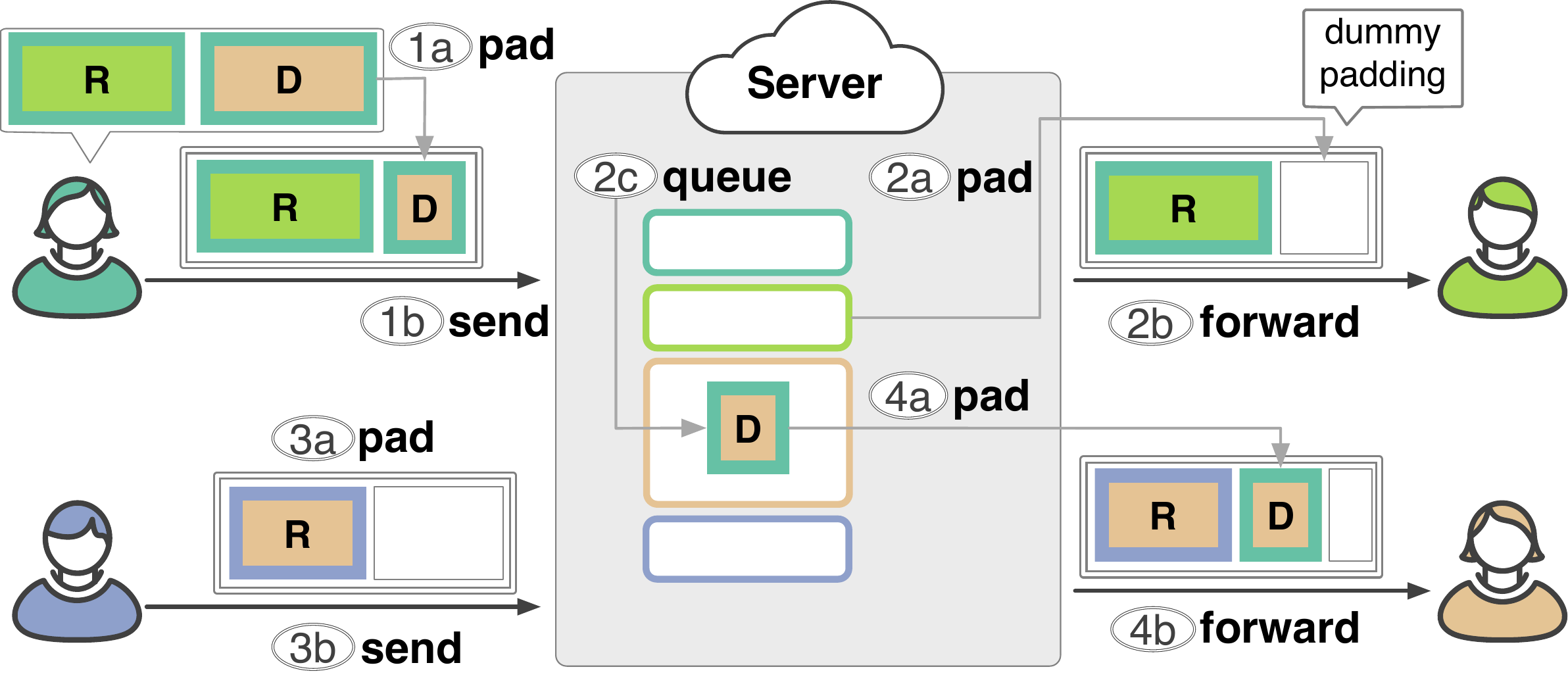}
    \caption{Diagram representation of the communication flow in \ourprotocol. 
    R and D denote regular and deniable communication, respectively. 
    Double lined boxes represent TLS tunneled traffic. 
    Odd steps (1 and 3) are performed by clients, even steps (2 and 4) are performed by the server.}\label{fig:our-protocol}
\end{figure}

\subsubsection{Deniable padding}\label{sec:padding}
The size of the deniable part of a network message is $\msgsize\padding$, where \msgsize\ is the length of the regular part, and \padding\ is a system-wide padding parameter set by the server. 
Both $\msgsize$ and $\padding$ are publicly known. 
Because of the strict size limit on the deniable part, deniable communication is chunked to fit the deniable part. 
If there is no deniable communication (or its length is less than $\msgsize\padding$), the deniable part is padded to always reach length~$\msgsize\padding$.

\subsubsection{Deniable buffers}\label{sec:buffers}
Each client keeps a deniable buffer to allow the user to queue new deniable messages at any time. 
Any time the user sends a regular message, $\msgsize\padding$ bytes of the oldest deniable message in the buffer are added as padding to the regular message.

The server keeps one deniable buffer per user. When receiving a message, the server extracts the regular part of the message, and creates a new message for the receiver and adds a deniable part -- either from the recipient's deniable buffer or dummy padding.
Note that depending on the implementation strategy, there may be subtle timing channels here. 
In particular, it may be desirable to handle the deniable parts of the incoming message only after the response has been processed. 
This is because differences in timing could leak information about the deniable part -- such as how many deniable actions were piggybacked.

\subsubsection{Changes to Signal}\label{sec:signal-changes}
A known (and often overseen) weakness of Signal is that if a user runs out of ephemeral keys, new sessions are initialized with less randomness by reusing the mid-term key instead of a one-time use key. Standard Signal mitigates this issue by letting users refill ephemeral keys at any point in time to avoid running out of keys.

However, if the key distribution center (KDC) were to fail to return a deniable ephemeral key for a user because they have run out of keys, it would leak information about the number of deniable sessions a user has. Therefore, the number of deniable ephemeral keys each user stores in the KDC must be secret to the adversary. To limit traffic between the KDC and the user, and still maintain the randomness for the generation of the master secret, \ourprotocol\ lets the KDC generate new deniable ephemeral keys on the behalf of the user. To keep the KDC and client in sync, the client provides a seed for the KDC to be used as input for a deterministic key generator upon user registration. The KDC keeps a counter for how many times the key generator has been used, and the value of the counter is sent to the corresponding client with each deniable message.
We stress that this change to Signal still means that the deniable messages will be end-to-end encrypted between sender and recipient -- the server will at most have access to one of the three keys (the ephemeral one) used to generate the master secret.

\subsubsection{Supported deniable actions}\label{sec:functionality}

In \ourprotocol\ we support all actions needed to implement the Signal protocol's session initialization and double ratchet mechanism. However, we do not support all functionality that the Signal IM app supports. 
For example, we do not support group chats and video calls. We have intentionally chosen not to support specific functionality to prevent adversaries from learning about users' deniable behavior. 
First, we do not support read receipts for messages, since they leak to an adversary if or when a user receives a  deniable message. 
Second, we support users blocking other users, but with the twist that blocked users are not informed that they have been blocked. 
An adversary that could learn that they have been blocked can for example flood a user with messages to provoke the user into blocking them, and the adversary can then use the time of being blocked to infer that a user has received their deniable messages, which also leaks that the user's deniable buffer has been drained.

In order to use \ourprotocol, each user needs to upload Signal keys and a seed for the key generator to the KDC through registering. 
We support the following \emph{deniable} Signal actions:

\ourparagraph{Key exchanges} 
Users can send key requests to initiate new Signal sessions, and the server responds with a key response containing the user's public identity key, mid-term key, and crucially, always an ephemeral key unlike in standard Signal where the KDC may run out of ephemeral keys. While the KDC always provides ephemeral keys as part of the key material, users can also issue key refills.

\ourparagraph{User message} 
Signal messages containing ciphertext and Signal headers to support ratcheting.

\ourparagraph{Block request} 
Enforced by the server by silently dropping messages by blocked users instead of buffering them.

\section{Formal analysis}\label{sec:formal-model}
This section presents the security and privacy analysis of \denimonsignal. 
In this formal privacy analysis we abstract away the details of the internal state of the unmodified version of Signal, which allows for a cleaner modeling using an abstract centralized protocol. 
Due to space, the full formal model and accompanying proofs are available \iffullversion{in \Cref{app:formal-model,app:indistinguishability,app:main-proof}}.\else{online\footnote{\url{https://www.dropbox.com/sh/90ufh9q3q74qoks/AAB6a9vnYtKTqCZrtM_Th71Oa?dl=0}}.}\fi

\subsection{Cryptographic guarantees}\label{sec:crypto-guarantees}
From the cryptographic point of view, \ourprotocol\ is an instance of Signal. Since our protocol does not alter the mechanisms employed to derive key material, encrypt, and decrypt messages, it trivially satisfies the cryptographic guarantees offered by the Signal protocol. 
More formally, since the security analysis of Cohn-Gordon et al.~\cite{cohn2020formal} rules out ephemeral key refill issues, it applies to \ourprotocol\ as well. This is straightforward for regular traffic, since \ourprotocol ephemeral keys are refilled as in Signal for regular messages. For deniable ones we fall back to use a secret seed shared between a user and the KDC. Employing a secure pseudo-random generator (PRG) to derive deniable ephemeral keys from the seed brings us to the setting of Cohn-Gordon et al.'s security analysis, since we assume the server to be honest.

\subsection{Protecting deniable behavior}\label{sec:metadata-privacy}
To ensure privacy against the network level \attacker\ described in \Cref{sec:adv-model}, we need to secure the two channels: the network, and the shared state.
Ultimately, an \attacker\ will examine network messages to try to distinguish dummy padding from deniable content, and observe the traffic patterns as well as trying to access shared states to infer something about a target user's deniable behavior.

\ourparagraph{Making deniable traffic indistinguishable}
Part of securing the network relies on guaranteeing that the \attacker\ cannot fingerprint the deniable part of network messages -- for this we need a TLS-like connection between each client and the server. 
TLS, however, is not sufficient, since the \attacker\ can participate in the protocol themselves, and it is important they do not infer anything about deniable behavior of other users through such participation.

\ourparagraph{Protecting traffic patterns and shared states}
We need to prove that a user's deniable behavior, which is reflected by the deniable protocol's internal state both client-side and server-side, does not leak into the regular protocol's state and becomes visible on the network or in the server's state.

The key technical insight of our approach is that by developing a fine-grained formal model of the deniable protocol and the tunnel, we can formulate the privacy problem as a form of noninterference~\cite{noninterference} property, where the confidential input to the system is users' deniable behavior. This allows us to leverage state-of-the-art formal machinery for information flow to precisely characterize the guarantees of the deniable protocol.
The main result of our analysis is crystallized at the end of this section as \Cref{thm:denim:privacy}.

The model formalizes both client and server behavior in \ourprotocol, but abstracts away the details of the tunnel that constitutes the regular part of a network message. 
For simplicity, we model the message forwarding server and the KDC as one `server' network node. To model Signal's double ratchet mechanism as part of our deniable protocol, we introduce a notion of \emph{abstract ratchets}, drawing on the techniques from the literature on symbolic cryptography~\cite{abadi1999secrecy}. Abstract ratchets turn out to be a particularly fitting technical gadget to reason about important privacy guarantees of 
\ourprotocol, without diving into probabilistic modeling. 
The rest of the section highlights some of the aspects of the formal model. 

\subsubsection{Communication model}
Our communication model uses the notion of upstream and downstream messages. Intuitively,
upstream events are the events that flow in the direction of the server, e.g., sending a message, or
registering a user. Downstream events are the events that flow from the server to a client. 
Each message consists of the client node designation $\nodemeta$, the host protocol payload $\hostpayloadmeta$ that we abstract away from, and a \ourprotocol event~$\eventmeta$. Additionally, downstream messages include the counter $\countermeta$ for synchronizing clients key generation with server-generated keys introduced in \Cref{sec:signal-changes}. 
$$
\msgmeta  ::=
\denimup{\nodemeta}{\hostpayloadmeta}{\eventmeta} \mid \denimdown{\nodemeta}{\hostpayloadmeta}{\eventmeta}{\countermeta}
$$

\ourprotocol events correspond to the supported deniable actions (cf \Cref{sec:denim-details}) and the corresponding server responses. 
The events contain information about the originator and the potential destination, as well as associated information, such as symbolic keys $\keymeta$ and symbolic ratchet tokens $\tokenmeta$. Here, we assume that keys are elements of the abstract key space~$\keysetmeta$. 

Ratchet tokens are sets of the form:
$$\token{\nodemeta_1}{\keymeta_1}{\nodemeta_2}{\keymeta_2}{\xaxismeta}{\yaxismeta}$$ and correspond to an active session between nodes $\nodemeta_1$ and $\nodemeta_2$, using $\keymeta_1$ and $\keymeta_2$ that are used for initiating the session. Coordinates $(\xaxismeta, \yaxismeta)$ correspond 
to the Signal message coordinates as described in \Cref{sec:signal-overview}.

Without loss of generality, we assume that all deniable events, including dummy padding, are of the same length. This allow us to omit the actual payload from the events grammar, and instead rely on payload confidentiality that we get from Signal. The following grammar describes \ourprotocol events:
\begin{align*}
\eventmeta  ::= &\ \evtsend{\nodemeta}{\nodemeta}{\tokenmeta} 
                \mid \evtfwd{\nodemeta}{\nodemeta}{\tokenmeta} 
              \\ \mid &\ \evtrefill{\nodemeta}{\keymeta} 
              \mid \evtkreq{\nodemeta}{\nodemeta} 
              \mid \evtkresp{\nodemeta}{\keymeta}{\nodemeta}\\
              \mid &\ \evtblock{\nodemeta}{\nodemeta} 
              \mid \dummymeta
\end{align*}
Here, the event $\evtsend{\nodemeta_1}{\nodemeta_2}{\tokenmeta}$ corresponds to the upstream event of sending a deniable message from $\nodemeta_1$ to $\nodemeta_2$. Once received by the server, the server forwards it to the destination as $\evtfwd{\nodemeta_1}{\nodemeta_2}{\tokenmeta}$. 

\begin{table}
\begin{center}
\begin{tabularx}{\linewidth}{llllX}
    \toprule
     \eventmeta & $\mathrm{dir}$ & $\mathrm{snd}$  &  $\mathrm{rcv}$& $\mathrm{kind}$ \\ \midrule
     \evtsend{\nodemeta_1}{\nodemeta_2}{\tokenmeta} & \upstreamevent  & $\nodemeta_1$  & $\nodemeta_2$ & \eventtypesend{\nodemeta_2} \\
     \evtfwd{\nodemeta_1}{\nodemeta_2}{\tokenmeta} & \quad \downstreamevent  & $\nodemeta_1$ & $\nodemeta_2$ & \\
     \evtrefill{\nodemeta}{\keymeta} & \upstreamevent & $\nodemeta$ &  & \eventtyperefill\\ 
     \evtkreq{\nodemeta_1}{\nodemeta_2} & \upstreamevent & $\nodemeta_2$ & & \eventtypekreq{\nodemeta_1}\\ 
     \evtkresp{\nodemeta_1}{\keymeta}{\nodemeta_2} & \quad \downstreamevent & & $\nodemeta_2$ & \\ 
     \evtblock{\nodemeta_1}{\nodemeta_2} & \upstreamevent & $\nodemeta_2 $ &  & 
     \eventtypeblock{\nodemeta_1} \\ 
    \dummymeta & & & & \eventtypedummy \\
    \bottomrule
\end{tabularx}
\caption{Auxiliary functions on unobservable events.\label{fig:aux:unobservable:event:fns}}
\end{center}
\end{table}

Based on the event grammar, we define a few auxiliary functions.

\begin{definition}[Upstream and downstream events; event sender and receiver]
Given an event $\eventmeta$ we define \emph{event direction}, denoted as $\eventdirection{\eventmeta}$, and \emph{event sender and receiver}, denoted as $\eventsender\eventmeta$ and $\eventreceiver\eventmeta$, respectively, as per \Cref{fig:aux:unobservable:event:fns}.
\end{definition}

\noindent
We lift these functions to messages, so we write, e.g., \eventdirection{\msgmeta} to get the direction of message \msgmeta.

Next, we introduce the notion of traces, and their local projections. 
\begin{definition}[\ourprotocol trace]
A \emph{\ourprotocol trace}, denoted as \tracemeta, is a sequence of \ourprotocol messages
$\msgmeta_1, \dots, \msgmeta_n$. Empty traces are denoted as $\emptytrace$.
\end{definition}

\begin{definition}[Local trace projection]
Given a trace $\tracemeta$, and a node $\nodemeta$, define \emph{local trace projection}, written \traceprojection{\tracemeta}{\nodemeta}, to be the subtrace of $\tracemeta$ consisting only of messages that are local to $\nodemeta$. It is defined inductively as $\traceprojection{\emptytrace}{\nodemeta} = \emptytrace$, and 
$$
\traceprojection{\msgmeta \tracecons \tracemeta}{\nodemeta} = 
\begin{cases}
\msgmeta \tracecons \traceprojection{\tracemeta}{\nodemeta}  & 
    \text{if~}  \eventdirection{\msgmeta} = \upstreamevent \land~\eventsender{\msgmeta} = \nodemeta\\ 
       & \lor~ \eventdirection{\msgmeta} = \downstreamevent \land~ \eventreceiver{\msgmeta} = \nodemeta \\
\traceprojection{\tracemeta}{\nodemeta} & \text{otherwise}
\end{cases}
$$
\end{definition}

To model users' deniable behavior, we introduce the notion of
strategies~\cite{strategies:cscfw06}. A strategy $\strategymeta$ is a function that takes a user-local view of
the trace and decides the next upstream event. These decisions eschew low-level details of ratcheting, instead only providing information about the kind of the event $\eventtypemeta$ given by the following grammar. 
$$
\eventtypemeta ::= 
    \eventtypesend \nodemeta  
    \mid \eventtyperefill  \mid \eventtypekreq{\nodemeta} \mid \eventtypeblock \nodemeta \mid \eventtypedummy
$$
\noindent 
Event kinds and events are related by function $\eventkind\eventmeta$, defined in \Cref{fig:aux:unobservable:event:fns}.
Next, we define a notion of strategy validity, which allows us to qualify user behavior. 
\begin{definition}[Strategy send validity]\label{def:strategy:validity}
Given a set of adversary nodes $\advnodesmeta$, and a strategy function $\strategymeta$ that runs on node $\nodemeta$, 
say that this strategy is \emph{send-valid w.r.t. $\advnodesmeta$},  
if the following conditions hold for all traces $\tracemeta$:
\begin{description}
\item[Send well-formedness] ~
if $\strategymeta(\tracemeta) = \eventtypesend \nodemeta_{\mathit{dest}}$ then $\tracemeta$
must contain message $\msgmeta$ such that either
\begin{itemize}
\item $\msgmeta = 
    \denimdown{\nodemeta}
              {\hostpayloadmeta}
              {\evtkresp{\nodemeta_{\mathit{dest}}}{\keymeta}{\nodemeta} }
              {\countermeta}
             $, or 
\item $\msgmeta = \denimdown{\nodemeta} {\hostpayloadmeta} { \evtfwd{\nodemeta_\mathit{dest}}{\nodemeta}{\tokenmeta}}{\countermeta}$
\end{itemize}

\item[\ourprotocol send validity] 
if $\strategymeta(\tracemeta) = \eventtypesend \nodemeta_{\mathit{dest}}$ or  
$\strategymeta(\tracemeta) = \eventtypekreq \nodemeta_{\mathit{dest}}$ for some node $\nodemeta_{\mathit{dest}}$, then $\nodemeta_{\mathit{dest}} \notin \advnodesmeta$.

\end{description}

\end{definition}

Send well-formedness is a technical requirement that ensures that the strategy follows the basic mechanics of the Signal protocol, since DenIM is a variant of Signal. In order to send a message to $\nodemeta_{\mathit{dest}}$, we either need to have obtained the key material from the server to initiate the session, or $\nodemeta_{\mathit{dest}}$ needs to have initiated the session already.  We note that there is no loss of generality in restricting the adversary to Signal messages in our model. All malformed deniable messages to the server will be dropped (in that way, they are morally equivalent to dummy messages); all malformed messages to the client will be dropped following the Signal rules.

\ourprotocol send validity states that non-malicious nodes do not 
communicate with the adversary. This is a critical constraint that is important for the \ourprotocol privacy theorem. The following example demonstrates an attack that is possible if \ourprotocol send validity does not hold.

\ourparagraph{Example attack}
Consider a system with three users: Alice, Bob, and the adversary, Eve. We assume that Eve observes all the network traffic, including messages from/to Alice/Bob via the server. Consider the trace consisting of the following messages.
\begin{enumerate}[noitemsep,nolistsep]
    \item Alice sends a message to the server  $\msgmeta_1 = \denimup{{\mathrm{Alice}}}{\hostpayloadmeta_1}{\eventmeta_1}$ with some host payload $\hostpayloadmeta_1$. Eve observes $\msgmeta_1$, but does not know whether $\eventmeta_1$ is a deniable message for Bob or not. If $\eventmeta_1$ is for Bob, it would mean that $\eventmeta_1$ must be queued on the server.
\item  Eve requests Bob's key material from the server:
    $\msgmeta_2 = \denimup{\mathrm{Eve}} {\hostpayloadmeta_2} {\evtkreq{\mathrm{Bob}}{\mathrm{Eve}}}$.
\item The server responds to Eve with Bob's key $\keymeta_1$ via message
         $\msgmeta_3 = \denimdown{\mathrm{Eve}} 
                                 {\hostpayloadmeta_3}
                                 {\evtkresp{\mathrm{Bob}}{\keymeta_1}{\mathrm{Eve}}}
                                 {\countermeta_1}$.
\item Eve uses $\keymeta_1$ to initiate a session with Bob. Eve generates their own key $\keymeta_2$ and construct a symbolic ratchet token
 $\tokenmeta_1 = \token{\mathrm{Eve}}{ \keymeta_2 }{ \mathrm{Bob} }{ \keymeta_1 }{ 0}{ 0 }$. Here, $(0,0)$ are the Signal coordinates of the message. 
 Eve constructs a deniable message $\eventmeta_2 = \evtsend{\mathrm{Eve}}{\mathrm{Bob}}{\tokenmeta_1}$, and sends message 
  $\msgmeta_4 = \denimup{\mathrm{Eve}} {\hostpayloadmeta_4} {\eventmeta_2}$. Upon receiving this message, the server must queue $\eventmeta_2$. 

\item The server sends a message 
      $\msgmeta_4 = \denimdown{\mathrm{Bob}}{\hostpayloadmeta_5} {\eventmeta_3}{\countermeta_2}$ to Bob. Here $\eventmeta_3$ is either $\eventmeta_1$, in case it was not dummy, or $\eventmeta_2$. 

\end{enumerate}
At this point, suppose that $\eventmeta_1 $ is dummy, and Bob receives Eve's $\eventmeta_2$ message. If Bob replies to it, they must construct a token 
$\tokenmeta_2 = \token{\mathrm{Eve}}{ \keymeta_2 }{ \mathrm{Bob} }{ \keymeta_1 }{ 1}{ 0 }$, Note the change in coordinate $1$ here. When Eve receives this message they know that Bob has received $\eventmeta_2$, because of the change in ratchet coordinates! In particular, Bob's message cannot be explained by Bob reaching out to Eve on their own. Additionally, receiving $\eventmeta_2$, when there is only one downstream message to Bob, also means that Bob did not receive $\eventmeta_1$, which leaks Alice's deniable behavior.

This example shows the importance of the \ourprotocol send validity requirement. A similar example can be constructed for allowing users to request the adversary's keys. Note that blocking an adversary is allowed by \Cref{def:strategy:validity}. The reason is that unlike sending a message to the adversary or requesting their key, the server does not propagate these events to the blocked users.

Note that our provable methodology means that \ourprotocol\ is 
secure against \emph{all} attacks that fall within the threat model, not just the example above. In fact, the example above has been discovered during the proof process.

A final technical aspect of strategy validity is that we require strategies to be \emph{deterministic w.r.t. formal randomness}. 
Intuitively, strategies may differentiate among keys, but cannot depend on the actual bit representation of the keys.
This reflects the probabilistic nature of generated keys, and is a natural requirement in symbolic cryptographic models.

\subsubsection{System state}
At the top level, our system is represented as network configurations of the form \systemconfig{\serverstatemeta}{\Userstatemeta}{\tracemeta}, 
where $\serverstatemeta$ is the server state, $\Userstatemeta$ is a collection of user states of the 
form $\userstratconfig{\strategymeta}{\usersignalstatemeta }$, where $\strategymeta$ is the user deniable strategy, and $\usersignalstatemeta$ is the
user's local Signal state that contains the necessary information for session bookkeeping. Finally, $\tracemeta$ is the global trace
produced so far. 

We formally define user and server state, respectively, as follows.
\begin{definition}[User local Signal state]
A user state $\usersignalstatemeta$ is a tuple \userconfig{\nodemeta}{\keyselfmeta}{\keyothersmeta}{\tokenstoremeta}{\seedmeta}{\countermeta}{\keygencounter}, where
\nodemeta\ is the identity of the user,
\keyselfmeta\ and \keyothersmeta\ are sets containing the user's own keys or keys of other users paired with a state of the form \keytuple{\keymeta}{\keyfresh} or \keytuple{\keymeta}{\keyused},
\tokenstoremeta\ is the abstraction of a ratchets mapped by receiving users to (\firstindexmeta, \{$\keymeta_1, \keymeta_2$\}, $\indexstoremeta$) containing the starting index assigned to the user, the initiating keys, and the observed message indices $\indexstoremeta$,
\seedmeta\ is a seed for the deterministic random number generator,
\countermeta\ is a formal randomness counter for the server-side key generation, and $\keygencounter$ is the formal randomness counter for the client-side key generation.
\end{definition}

\begin{definition}[Server state]
A server state is represented by a tuple 
\serverconfig{\hoststatemeta}{\servermessagedownqueue}{\deniablestatemeta} 
where \hoststatemeta\ represents the state of an arbitrary host protocol, 
\servermessagedownqueue\ is a list of outgoing regular messages, 
and \deniablestatemeta\ represents the deniable state. The deniable state is a mapping of a user to a tuple of the form \deniablestatetuple{\seedmeta}{\countermeta}{\keystoremeta}{\blockliststoremeta}{\servereventdownqueue}, where \seedmeta\ represents a seed for a deterministic random number generator, \countermeta\ a counter, \keystoremeta\ a set of keys, \blockliststoremeta\ a set of blocked users, and \servereventdownqueue\ a list of \ourprotocol\ payloads.
\end{definition}

The notion of validity is lifted to user configurations: strategies of non-adversarial nodes must be valid, written $\validstrategy{\Userstatemeta}{\advnodesmeta}$. We assume that all users are initially registered on the server with their respective secret seeds.

\begin{figure}
\begin{framed}
\begin{mathpar}
\inferrule[Net-Global]{ 
\Userstatemeta = \userstatemeta_1 \dots \userstatemeta_j \dots \userstatemeta_n\\
{\userstatemeta_j}
\userauxstep{\tracemeta}{\msgmeta}
{\userstatemeta_j}'\\
\Userstatemeta' = \userstatemeta_1 \dots {\userstatemeta_j}' \dots \userstatemeta_n\\
\serverstatemeta
\serverauxstep{}{\msgmeta}
\serverstatemeta'\\
}{ 
\systemconfig{\serverstatemeta}{\Userstatemeta}{\tracemeta}
\networkstep 
\systemconfig{\serverstatemeta'}{\Userstatemeta'}{\tracemeta\tracecons\msgmeta}
}\\
\end{mathpar}
\end{framed}
\caption{Network transitions\label{fig:network-global-transitions-main-paper}}
\end{figure}

\begin{figure}
\begin{framed}
\begin{mathpar}
\inferrule[Aux-Upstream-User-Event]{
\usersignalstatemeta = \userconfig{\useridmeta}{\keyselfmeta}{\keyothersmeta}{\tokenstoremeta}{\seedmeta}{\countermeta}{\keygencounter}\\ 
{\usersignalstatemeta}
\userdenimstep{}{\eventmeta}
{\usersignalstatemeta}'\\
\msgmeta = \denimup{\nodemeta}{\hostpayloadmeta}{\eventmeta} \\
\strategymeta(\traceprojection{\tracemeta}{\nodemeta})  = \eventkind{\eventmeta} 
}{
\userstratconfig{\strategymeta}{\usersignalstatemeta}
\userauxstep{\tracemeta}{\msgmeta}
\userstratconfig{\strategymeta}{\usersignalstatemeta'}\\
}\and 
\inferrule[Aux-Upstream-User-\dummymeta]{
\usersignalstatemeta = \userconfig{\useridmeta}{\keyselfmeta}{\keyothersmeta}{\tokenstoremeta}{\seedmeta}{\countermeta}{\keygencounter}\\ 
\msgmeta = \denimup{\nodemeta}{\hostpayloadmeta}{\dummymeta} \\
\strategymeta(\traceprojection{\tracemeta}{\nodemeta}) = \eventtypedummy
}{ 
\userstratconfig{\strategymeta}{\usersignalstatemeta}
\userauxstep{\tracemeta}{\msgmeta}
\userstratconfig{\strategymeta}{\usersignalstatemeta}\\
}
\end{mathpar}
\end{framed}
\caption{Auxiliary user transitions: selected rules\label{fig:aux-user-transitions:main-paper}}
\end{figure}

\subsubsection{System transitions}

\Cref{fig:network-global-transitions-main-paper} captures 
the top-level interaction between users and the server.  It shows how the system is updated when a user sends a message, $\msgmeta$.
\Cref{fig:aux-user-transitions:main-paper} presents selected rules 
for the upstream user state transitions $\userstratconfig{\strategymeta}{\usersignalstatemeta}
\userauxstep{\tracemeta}{\msgmeta}
\userstratconfig{\strategymeta}{\usersignalstatemeta'}
$.
Here, $\tracemeta$ is the trace of the system so-far, and $\msgmeta$
is the new upstream message. The host protocol payload $\hostpayloadmeta$
is chosen non-deterministically, while the \ourprotocol event is
constrained by the strategy function $\strategymeta$. The subtlety of these
rules is that the strategy determines only the kind of the next deniable message, but not their exact content, because the latter depends on the ratchet coordinates. There are two advantages of having the strategies define only the kind of the event and not its full content. First, this keeps the notion of strategy well-formedness simple; it would otherwise have to rule out nonsensical sequences of ratchet coordinates. Second, as a modeling device, it is also conceptually more truthful to capturing the user intent, e.g., the one of sending a message rather than sending a message with particular ratchet indices. Finally, note that we use two rules for the upstream messages. When the user strategy returns dummy, $\dummymeta$, we do not update the user local Signal state. Note the clause ${\usersignalstatemeta}\userdenimstep{}{\eventmeta} {\usersignalstatemeta}'$ in the rule (Aux-Upstream-User-Event) and its lack in the rule (Aux-Upstream-User-\!$\dummymeta$).

Our main theorem states that all valid DenIM strategies are possible. It is formulated as a noninterference theorem on strategies. We write $\abstractloweq{}{}{\advnodesmeta}{}$ when two configurations or traces are indistinguishable by the adversary. 

\begin{theorem}[\ourprotocol privacy]\label{thm:denim:privacy}
Consider a set of adversary nodes $\advnodesmeta$, and two initial indistinguishable configurations
$
\abstractloweq{\systemconfig{\serverstatemeta_1}{\Userstatemeta_1}{\emptytrace}}{\systemconfig{\serverstatemeta_2}{\Userstatemeta_2}{\emptytrace}}{\advnodesmeta}{\initeqmarker}
$, with valid user strategies, that is $\validstrategy{\Userstatemeta_i}{\advnodesmeta}, i=1,2$.
If 
$
\systemconfig{\serverstatemeta_1}{\Userstatemeta_1}{\emptytrace} 
\networkstepmany 
\systemconfig{\serverstatemeta'_1}{\Userstatemeta'_1}{\tracemeta_1}
$
then 
$
\systemconfig{\serverstatemeta_2}{\Userstatemeta_2}{\emptytrace} 
\networkstepmany 
\systemconfig{\serverstatemeta'_2}{\Userstatemeta'_2}{\tracemeta_2}
$,
such that $\abstractloweq{\tracemeta_1}{\tracemeta_2}{\advnodesmeta}{}$.
\end{theorem}

\begin{proof}
By induction on the length of $\tracemeta_1$ using \Cref{lemma:unwinding} (see \Cref{app:formal-model}).
 
\end{proof}

\section{Empirical evaluation}\label{sec:evaluation}
To evaluate the feasibility and performance of \ourprotocol, we have created a proof-of-concept implementation of \denimonsignal\ and designed experiments to measure the system's behavior when running on a real network.

\subsection{Implementation}\label{sec:implementation}
Our \ourprotocol implementation runs on NodeJS. 
It amounts to 3838 lines of TypeScript code\footnote{\url{https://www.dropbox.com/sh/cvmo7op81wf8jbj/AABx0buTpJj09Q1CDyPsim2Ga?dl=0}}, using the Signal app's open-source implementation~\footnote{\url{https://github.com/signalapp/libsignal}} of the Signal protocol and their TypeScript bindings (developed for the official Signal Desktop Client). 
We implement three types of network entities: a dispatcher server that orchestrates the experiments and injects code in the clients to simulate different user behaviors, a \ourprotocol\ server which handles messages and also acts as a KDC, and \ourprotocol\ clients. 

\subsubsection{Network messages}\label{sec:message-format}

Crucial to achieving \ourprotocol's unobservability of deniable traffic is producing network traffic that does two things: 1) ensures that dummy padding is indistinguishable from encrypted deniable payloads, and 2) ensures that any given padding size can be precisely achieved. 
To achieve 1), we use TLS sockets for client-server communication, and to achieve 2) we have designed own message formats to ensure that padding serializes to the right size.

To represent and serialize objects, we use Google's protocol buffers~\cite{google_developers_encoding_2022} -- a language and platform-agnostic mechanism for serializing structured data that is also used by the official Signal implementation. All communication is packaged within a \texttt{DenimMessage} structure (see \Cref{code:denim-message}) before being serialized and sent over an encrypted TLS connection. Regular communication is stored within \texttt{RegularPayload} structure, the ratio of deniable padding to use is communicated by the server using \texttt{q}, and \texttt{c} is used by the server to keep the client's and server's deniable ephemeral key generator in sync. The deniable communication is chopped up to fit within the allotted deniable padding, and stored in the field \texttt{chunks} on the \texttt{DenimChunk} structure (\Cref{code:denim-chunk}). Since the deniable padding needs to be a fixed size and Google's protocol buffers use varint encoding, we use two additional ballast fields to be able to vary the length of the serialized object by one byte.

\begin{lstlisting}[language=Protobuf,label=code:denim-message,caption={Network message serialization format.}]
message DenimMessage {
 required RegularPayload payload
 optional double q  //server -> client
 optional int32 c //server -> client
 repeated DenimChunk chunks
 required int32 ballast;
 optional int32 extra_ballast;
}
\end{lstlisting}

\begin{minipage}{\linewidth}
\begin{lstlisting}[language=Protobuf,label=code:denim-chunk,caption={Chunk serialization format.}]
message DenimChunk {
	required bytes chunk
	required int32 flags
}
\end{lstlisting}
\end{minipage}

\subsubsection{Randomized key labels}
In the implementation of Signal, each ephemeral key is labeled with an identifier to allow for quick matching of keys. The number of sessions a user has is not secret in regular Signal, so this identifier may be assigned sequentially. In \ourprotocol, however, assigning identifiers sequentially would leak to anyone requesting ephemeral keys from the KDC how many ephemeral keys the client has generated. To mitigate this leakage in \ourprotocol, we assign random identifiers to deniable ephemeral keys. Since the KDC can generate deniable ephemeral keys as well, the client and KDC stays in sync by providing a seed to generate a deterministic sequence of identifiers when registering with the KDC. Whenever the KDC sends a counter value higher than the client's current counter value, the client generates the corresponding amount of deniable ephemeral keys and identifiers using the two deterministic generators. To avoid identifier collisions, the output space is segmented into identifiers that can be used solely by the client, and solely by the server.

\subsection{Experimental setup and design}\label{sec:experiment-setup}
Each experiment run consists of one \ourprotocol\ server running on a dedicated machine, multiple clients evenly spread across four machines, and one dispatcher server. The \ourprotocol\ server is a remote machine with a dedicated 8 core 2.50GHz CPU and 64GB RAM running Ubuntu, and the clients are virtual machines running Ubuntu on 4 core 2.49GHz CPUs with 8GB RAM. The dispatcher server is a virtual machine running Ubuntu on 4 core 2.29GHz CPUs with 8GB RAM.

We have designed our experiments to evaluate the \ourprotocol's performance at high CPU loads, in the range of 80\%-90\% CPU utilization. Using the CPU utilization goal, we have tuned the number of clients to 20, and the client events to occur every tick of 20 ms. Every experiment run presented here is 60 seconds -- as we have seen no empirical changes in the system behavior in longer runs, up to 30 minutes. Each client establishes a TCP connection with the \ourprotocol\ server and performs the following steps:
\begin{enumerate}[noitemsep,nolistsep]
    \item Registers the user.
    \item Every 20 ms (tick), the clients does the following: \begin{enumerate}[noitemsep,nolistsep]
        \item sends a given amount of regular messages, to randomly chosen recipients.
        \item sends a given amount of deniable messages, to randomly chosen recipients.
    \end{enumerate}
\end{enumerate}
Each message consists of a 52 characters long delimited timestamp to measure latency, and a randomly chosen quote of mean length 57 characters, making messages 109 characters long on average.

To evaluate the performance impact of deniable traffic, we vary the variable \padding\ (experiment settings in \Cref{tab:experiment-settings}) that determines the amount of deniable traffic carried by regular traffic, and the proportion of deniable messages generated in relation to regular messages (experiment settings in \Cref{tab:experiment-settings-x}). We empirically establish that the size of a deniable message is approximately 1.2 the size of a regular message with the same plaintext, so when when increasing the ratio with 0.1 we vary \padding\ using steps of 0.12.

\begin{table}[htb]
    \centering
    \begin{tabularx}{\linewidth}{Xlll}
    \toprule
        \textbf{Id} &\textbf{\padding} & \textbf{Regular msgs/tick} & \textbf{Deniable msgs/tick}\\\midrule
        A1 & 0 & 10 & 0\\
        A2 & 0.12 & 10 & 1 \\
        A3 & 0.24 & 10 & 2\\
        A4 & 0.36 & 10 & 3\\
        A5 & 0.48 & 10& 4\\
        A6 & 0.6 & 10 & 5\\
        A7 & 0.72 & 10 & 6\\
        A8 & 0.84 & 10 & 7\\
        A9 & 0.96 & 10 & 8\\
        A10 & 1.08 & 10 & 9\\
        A11 & 1.2 & 10 & 10\\
        \bottomrule
    \end{tabularx}
    \caption{Experiment settings with \padding\ set in proportion to the regular and deniable messages ratio.}
    \label{tab:experiment-settings}
\end{table}

\begin{table}[htb]
    \centering
    \begin{tabularx}{\linewidth}{Xlll}
    \toprule
        \textbf{Id} &\padding & \textbf{Regular msgs/s} & \textbf{Deniable msgs/s}\\\midrule
        A6X1 & 0.6 & 10 & 1\\
        A6 & 0.6 & 10 & 5\\
        A6X2 & 0.6 & 10 & 10\\
        \bottomrule
    \end{tabularx}
    \caption{Experiment settings to capture behavior when \padding\ is not set in proportion to the regular and deniable messages ratio.}
    \label{tab:experiment-settings-x}
\end{table}

During the experiments, we measure CPU utilization, regular message throughput per second, and the length of the deniable buffers server-side. Client-side, we measure end-to-end latency for regular and deniable messages.

\subsection{Results}\label{sec:experiment-results}

The results for all settings from \Cref{tab:experiment-settings} (for all graphs see 
\iffullversion\Cref{app:experiment-graphs}
\else
online\footnote{\url{https://www.dropbox.com/sh/90ufh9q3q74qoks/AAB6a9vnYtKTqCZrtM_Th71Oa?dl=0}}\fi) show, as expected, that regular message throughput is higher for lower values of \padding, which means lower \padding\ gives lower regular message latency. We show the mean throughput for regular and deniable messages for each experiment setting in \Cref{tab:throughput-regular-deniable}.
For comparison, CoverDrop~\cite{ahmed-rengers_coverdrop_2022} empirically has a throughput of 833 (analogous to our deniable) messages per second.
Notice that at $\padding=1.2$ in setting A11, the throughput for regular and deniable traffic is similar which we expect due to the additional overhead in deniable messages compared to regular messages.

\begin{table}[htb]
    \centering
    \begin{tabularx}{\linewidth}{XXllXll}
    \toprule
        \textbf{Id} & \multicolumn{3}{c}{\textbf{Regular msgs}} & \multicolumn{3}{c}{\textbf{Deniable msgs}} \\ \cline{2-7}
         & s & per s & per min & s & per s & per min\\\midrule
        A1 & 0.014 & 7527 & 452k & - & 0 & 0 \\
        A2 & 0.021 & 7299 & 438k & 8.880 & 549 & 33k\\
        A3 & 0.018 & 7111 & 427k & 2.370 & 1352 & 81k\\
        A4 & 0.024 & 6888 & 413k & 0.366 & 2065 & 124k\\
        A5 & 0.026 & 6599 & 396k & 0.177 & 2638 & 159k\\
        A6 & 0.028 & 6544 & 393k & 0.140 & 3270 & 196k\\
        A7 & 0.030 & 6338 & 380k & 0.124 & 3801 & 228k\\
        A8 & 0.032 & 6163 & 370k & 0.0114 & 4312 & 259k\\
        A9 & 0.034 & 5998 & 360k & 0.0112 & 4796 & 288k\\
        A10 & 0.036 & 5955 & 357k & 0.111 & 5356 & 321k\\
        A11 & 0.038 & 5709 & 343k & 0.108 & 5706 & 342k\\
        \bottomrule
    \end{tabularx}
    \caption{The mean overhead for latency, throughput per second, and throughput per minute, for regular and deniable messages respectively}
    \label{tab:throughput-regular-deniable}
\end{table}

In \Cref{fig:graph-server-statistics} we show the server's CPU utilization and the regular message throughput over time for setting A6 where $\padding=0.6$ and the regular to deniable message ratio is 10:5.
The shape of the regular message throughput follows the CPU utilization, with an average of 6544 messages/s which is approximately 392k messages/min.
As we can see, the mean CPU utilization is 87\%, and the spikes in CPU utilization are correlated with the NodeJS major garbage collection events. 
Note that NodeJS's garbage collection is concurrent, whereas NodeJS's runtime is single-threaded; this can result in CPU loads greater than 100\% when garbage collection is running.
CPU load drops initially due to the clients waiting for key responses before they can encrypt messages.

\begin{figure}[htb]
     \centering
     \begin{subfigure}[b]{0.4\textwidth}
         \centering
         \includegraphics[width=\textwidth]{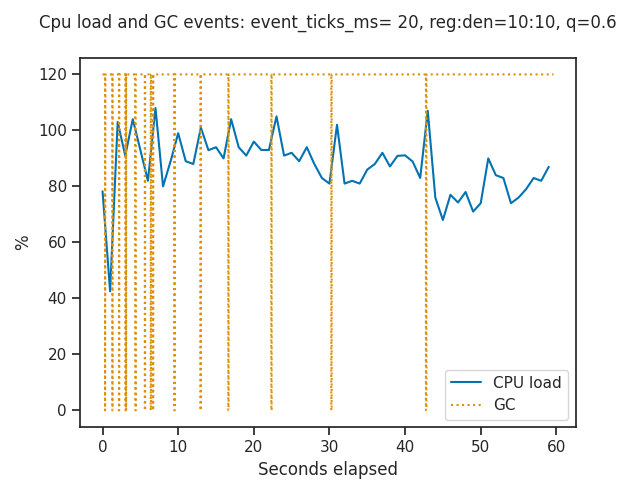}
         \caption{CPU load, note that spikes correlate with major garbage collection (GC) events.}
         \label{fig:graph-cpu}
     \end{subfigure}
     \hfill
     \begin{subfigure}[b]{0.4\textwidth}
         \centering
         \includegraphics[width=\textwidth]{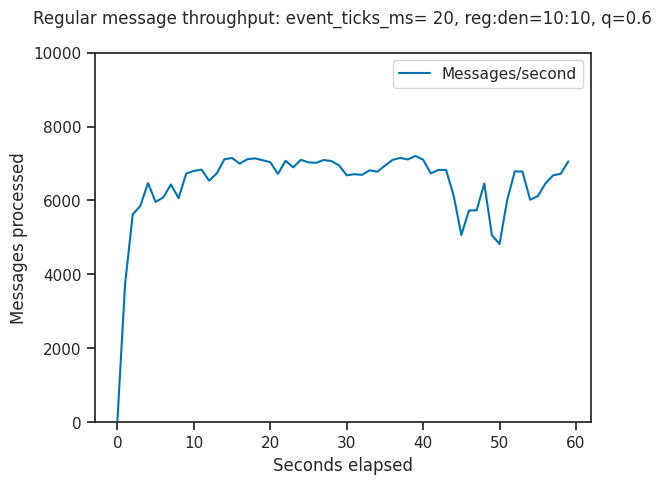}
         \caption{Regular messages processed over time.}
         \label{fig:graph-throughput}
     \end{subfigure}
\caption{Server statistics at $\padding=0.6$.}\label{fig:graph-server-statistics}
\end{figure}

In \Cref{fig:graph-regular-latency-zoom} we show a box plot of the latency for regular messages measured in seconds. The latency increases when the value of \padding\ increases, as the server spends more time chunking and reassembling deniable payloads. As a baseline, we have included A1 where $\padding=0$, which means no deniable traffic. The mean latency for A1 is 0.014s/message, and 0.038s/message for A11 where $\padding=1.2$. 

\begin{figure}[htb]
     \centering
     \centering
     \includegraphics[width=0.7\columnwidth]{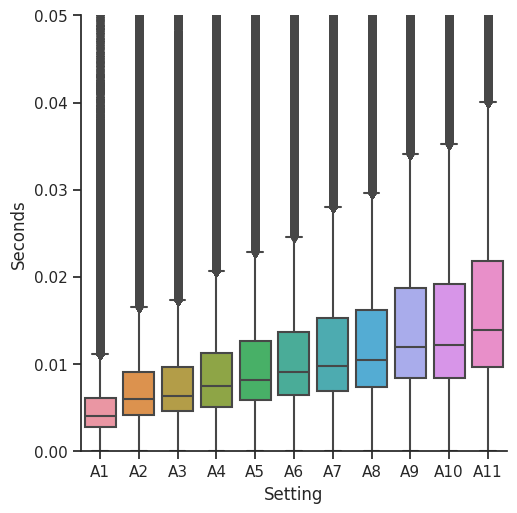}
\caption{Client to client message regular latency measure from $\padding=0$ to $\padding=1.2$ with step 0.12.}\label{fig:graph-regular-latency-zoom}
\end{figure}

For settings with a value of \padding\ that corresponds to 1.2 times the deniable to regular message ratio, e.g. $\padding=0.6$ when the ratio is 10:5, the deniable buffers get drained and filled at a similar rate after approximately 10 seconds\iffullversion~(see graphs in \Cref{app:experiment-graphs})\fi. When the deniable padding is a relatively small number (recall that the average message plaintext is 109 characters) in comparison to the size of the \ourprotocol\ headers, the deniable buffers can continue to grow on the server even when \padding\ is set in proportion to the deniable to regular message ratio. In our experiments, this happens at $\padding < 0.36$, i.e. for setting A2 and A3.

When the deniable buffers grow in size on the server, it contributes to large latency for deniable messages. We can see this for A2 and A3\iffullversion~in \Cref{fig:graph-deniable-latency} in \Cref{app:experiment-graphs}\fi, with a mean latency of 8.9s for A2, and 2.4s for A3. In \Cref{fig:graph-deniable-latency-zoom} we have adjusted the scale to capture the settings with $\padding \geq 0.36$, i.e. settings A4 to A11. For A4 to A7 we see latency decreasing, and remaining stable from A8 to A11 with a mean latency of 0.11s/message.

\begin{figure}[htb]
     \centering
         \centering
         \includegraphics[width=0.7\columnwidth]{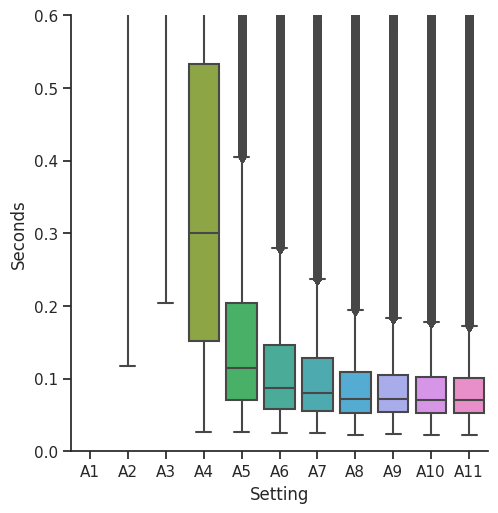}
\caption{Client to client message latency deniable measure from $\padding=0$ to $\padding=1.2$ with step 0.12. Notice change in scale.}\label{fig:graph-deniable-latency-zoom}
\end{figure}

Notice that each client has their own deniable buffer on the server, so clients' latency for receiving deniable messages depends on how many regular messages they receive in proportion to deniable messages, and how that proportion relates to the value of \padding. We use the settings from \Cref{tab:experiment-settings-x} and show in \Cref{fig:graph-buffers} that a higher proportion of deniable messages to regular messages than supported by \padding\ (10:10 which would be supported by $\padding=1.2$) results in growing deniable buffers, and increased latency for deniable messages. In the same graph, we show that a lower proportion (10:1 which would be supported by $\padding=0.12$) allows the server to drain the deniable buffer, and while it lowers latency, it does not increase the throughput of deniable messages since there are no deniable messages for the server to process. A similar fill and drain rate of the deniable buffers is achieved when \padding\ is set to reflect the difference in regular to deniable message ratio, in this case 10:5 using setting A6, and as previously shown with the experiment settings A1 to A11. 

\begin{figure}[htb]
    \centering
    \includegraphics[width=0.9\columnwidth]{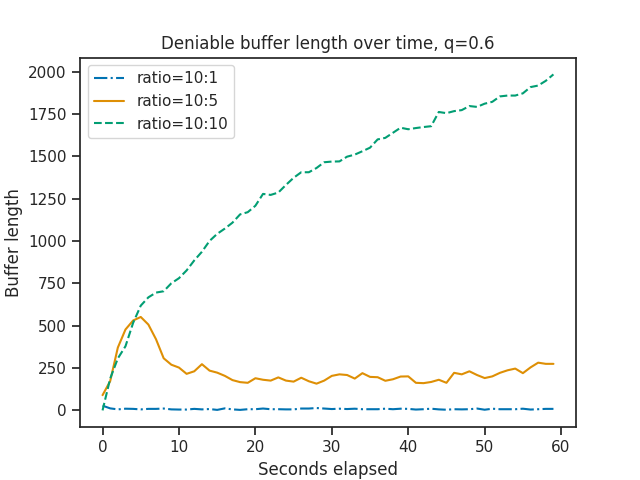}
    \caption{Length of deniable buffers when varying regular to deniable message ratio at $\padding=0.6$.}\label{fig:graph-buffers}
\end{figure}

\section{Limitations and future work}\label{sec:limitations}
This section discusses some current limitations of \ourprotocol and provides our perspective on how they may lifted.
We also elaborate on how the approach to designing \denimonsignal could be generalized to a methodology for designing data-aware tunneling protocols.

\ourparagraph{Latency of deniable messages and practical trade-offs}
A conscious design decision behind \denimonsignal is to prioritize strong privacy over low latency. 
This naturally affects quality of service, and is a design decision we have taken to reduce the performance overhead of introducing deniable messages to IM services.
A consequence of not prioritizing low latency is that a deniable message could remain undelivered, if the recipient has no incoming regular traffic. 
In other words, the currently chosen trade-off for \denimonsignal supports delivery of deniable messages for users that \emph{also} send and receive regular messages, and is not intended for users who \emph{only} want to send deniable messages.
We stress that users always are guaranteed privacy for their deniable messages in our system, whereas we provide no formal guarantees for utility.

It is possible to alter the trade-off and reduce latency by introducing additional bandwidth overhead.
For example, we envision that a practical deployment of \denimonsignal could feature a subscription feed, a group chat, or a Telegram-like channel, all of which would facilitate a steady stream of regular traffic to contribute to draining the deniable buffers server-side.
Along the same lines, a realistic client can periodically send out heartbeats or statistical information to the server that can be used to push deniable messages out of the server.

\ourparagraph{Security of the deniable message buffers}
The server deniable message buffer can turn into an attractive attack target. 
We note that the messages in the buffer are end-to-end encrypted, and only include the recipient information. 
To fully protect the state of the buffer, one can furthermore use trusted enclaves, as proposed by Ahmed-Rengers~et~al. in CoverDrop~\cite{ahmed-rengers_coverdrop_2022}.

\ourparagraph{Communicating with the \attacker}
While our current assumption of users not communicating with the \attacker\ aligns with the threat model of IM apps -- e.g., Signal and WhatsApp both warn users from communicating with untrusted parties rather than enforce the assumption -- future work may focus on lifting this assumption.
Because this will
break noninterference, the top-level security condition 
will need to be weakened. Here, we anticipate that 
either techniques from quantitative information flow~\cite{alvim_science_2020} or declassification~\cite{sabelfeld_dimensions_2005}
can provide useful security characterizations.

\ourparagraph{Utility guarantees}
Our current system provides formal guarantees for privacy, but not utility.
Future work could focus either on formally modeling user behavior to provide utility guarantees, or on extending the empirical evaluation to include real user data.
Acquiring real IM data is difficult since it is highly sensitive data that is not publicly available -- previous work such as Bahramali et al.~\cite{bahramali_practical_2020} have instead collected (but not released) data from alternate sources, namely public Telegram channels.
While such alternate data is possible to collect, although not from Signal as Signal does not support public channels, we expect public channel chats to significantly differ from our intended use case which is one-on-one chats.
In particular, channels allow a small group of admins broadcast messages to a large crowd, whereas group chats or one-on-one chats puts no restrictions on who can send messages. 
As such, the collection of quality data from IM usage would be a valuable, non trivial contribution for future work.

\ourparagraph{Parameter tuning}
The only parameter in our design that needs tuning in a real-world deployment is the padding, $\padding$. 
We stress that this parameter should be global and not be personalized, as individual choice of $\padding$ can be discriminating~\cite{bahramali_practical_2020}.

\ourparagraph{Generalizing the methodology}
While this paper focuses on building a specific system for IM, \denimonsignal, we anticipate the approach to translate into a generalized methodology to use IFC techniques for anonymous communication problems.

In particular, we expect that systems do not need to use an observable tunnel (in our case unmodified Signal) for noninterference to be a desirable property to prove to achieve metadata privacy -- we posit that noninterference between a deniable protocol and an anonymous communication protocol (for example DC-nets) would still be useful to show since our approach allows the tracking of all information flows from the deniable protocol rather than just the network layer.

Furthermore, we expect that an interesting direction for future work would be to investigate how similar the architecture, e.g., server structure, of the tunnel protocol needs to be to that of the deniable protocol.
\section{Related work}\label{sec:related-work}

\begin{table*}[tb]
    \centering
    \begin{tabularx}{\linewidth}{llllXX}
        \toprule
    \textbf{Protocol} & \textbf{Tunneling} & \textbf{Censorship resilience} &\textbf{Provable guarantees} & \textbf{Trust} & \textbf{Threat model}  \\
    \midrule

        DC-nets~\cite{chaum_dining_1988} & \xmark & Weak & Transport layer & Anytrust & GA \\
        Dissent~\cite{corrigan-gibbs_dissent_2010} & \xmark & Weak & Transport layer & Anytrust & GA \\
        Anonycaster~\cite{head_anonycaster_2012} & \xmark & Weak & Transport layer & Anytrust & GA \\
        Riffle~\cite{kwon_riffle_2016} & \xmark & Weak & Transport layer & Anytrust & GA \\ 
        Atom~\cite{kwon_atom_2017} & \xmark & Weak & Transport layer & Anytrust & GA \\ 
        Talek~\cite{cheng_talek_2020} & \xmark & Weak & Transport layer & Anytrust & GA \\

        Herd~\cite{le_blond_herd_2015} & \xmark & Weak & Transport layer & Chosen set & GP/LA \\ 

        Pynchon~\cite{sassaman_pynchon_2005}  & \xmark & Weak & Transport layer & Fraction & GA\\ 
        XRD~\cite{kwon_xrd_2020} & \xmark & Weak & Transport layer & Fraction & GA\\
        Express~\cite{eskandarian_express_2021} & \xmark& Weak & Transport layer & Fraction & GA\\ 

        $P^3$~\cite{kissner_private_2004} & \xmark & Weak & Transport layer & Honest-but-curious/Malicious & GA \\ 
        Loopix~\cite{piotrowska_loopix_2017} & \xmark & Weak & Transport layer & Honest-but-curious & GA \\ 
        
        Bitmessage~\cite{warren_bitmessage_2012}  & \xmark & Weak & Transport layer & Zero-trust & GA \\ 
        Pung~\cite{angel_unobservable_2016} & \xmark & Weak & Transport layer & Zero-trust & GA\\ 
        Riposte~\cite{corrigan-gibbs_riposte_2015} & \xmark & Weak & Transport layer &  Zero-trust*** & GA\\ 
        

        Verdict~\cite{corrigan-gibbs_proactively_2013} & \xmark & Weak & Transport layer* & Anytrust & GA \\
        Stadium~\cite{tyagi_stadium_2017} & \xmark & Weak &  Transport layer* & Anytrust & GA\\ 
        
        Vuvuzela~\cite{van_den_hooff_vuvuzela_2015} & \xmark & Weak &  Transport layer* & Fraction & GA \\
        Alpenhorn~\cite{lazar_alpenhorn_2016} & \xmark & Weak &  Transport layer* & Fraction & GA \\ 
        Karaoke~\cite{lazar_karaoke_2018} & \xmark & Weak &  Transport layer* & Fraction & GP \\ 
        Yodel~\cite{lazar_yodel_2019} & \xmark & Weak & Transport layer** & Fraction & GA\\ 
        
        Groove~\cite{barman_groove_2022} & \xmark & Weak & Transport layer* & Zero-trust & GA \\ 


        Mixminion~\cite{danezis_mixminion_2003} & \xmark & Weak & \noguarantee & Chosen path & GA\\ 
        HORNET~\cite{chen_hornet_2015} & \xmark & Weak & \noguarantee & Fraction & LA \\ 
        Tor~\cite{dingledine_tor_2004} & \xmark &  Weak & \noguarantee & One**** & GP \\


        \midrule
        \ourprotocol & \cmark & Strong & Application layer & Centralized & GA \\
        \midrule


        CoverDrop~\cite{ahmed-rengers_coverdrop_2022} & \cmark & Strong & \noguarantee & Centralized & GP \\ 
        
        Cirripede~\cite{houmansadr_cirripede_2011} & \cmark & Strong & \noguarantee & Proxies & LA \\
        Telex~\cite{wustrow_telex_2011} & \cmark & Strong & \noguarantee & Proxies & LA\\ 
        CensorSpoofer~\cite{wang_censorspoofer_2012} & \cmark & Strong & \noguarantee & Proxies & LA \\ 
        FreeWave~\cite{houmansadr_i_2013}  & \cmark & Strong & \noguarantee & Proxies & LA\\ 

        SkypeMorph~\cite{mohajeri_moghaddam_skypemorph_2012} & \cmark & Strong & \noguarantee & Proxy & LA \\
        IMProxy~\cite{bahramali_practical_2020} & \cmark & Strong & \noguarantee & Proxy & LA\\
        Protozoa~\cite{barradas_poking_2020} & \cmark & Strong & \noguarantee & Proxy & LA \\
        Camoufler\cite{sharma_camoufler_2021} & \cmark & Strong & \noguarantee & Proxy & LA  \\
        
        Balboa~\cite{rosen_balboa_2021} & \cmark & Strong & \noguarantee & Zero-trust & GA \\

    \bottomrule
    \end{tabularx}
    \caption{Comparison of related work. Footnotes: *via differential privacy, **with failure probability $10^{-8}$ per round, ***for privacy, all servers need to be trusted for availability, ****in chosen set. G=Global, L=Local, A=Active, P=Passive.}
    \label{tab:app-rw-comparison-full}
\end{table*}

\ourparagraph{Anonymous communication}
We review the related literature through the categorization in \Cref{tab:app-rw-comparison-full}.
This categorization is based on five dimensions.
First, \emph{tunneling}: ``does the approach tunnel traffic using an innocuous protocol?''.
Second, \emph{censorship resilience}: ``is the design intended to be difficult to detect and block?''.
Third, \emph{provable guarantees}: ``is privacy formally proven, and in that case, on what level of the network stack are the security guarantees?''.
Fourth, \emph{trust}: ``what part of the network is considered trusted?''.
And last, \emph{threat model}: ``what are the capabilities of the adversary?''.

Related work fall into four main clusters.
The first cluster provide provable guarantees on the transport layer, achieved mainly using DC-nets and mixes.
None of these approaches model the application layer, and therefore the privacy guarantees do not extend to e.g., shared protocol states.

The second cluster is similar to the first, but instead give probabilistic privacy guarantees, which allows them to achieve either lower latency or bandwidth but not both. 
This limitation is in line with the anonymity trilemma of Das~et~al.~\cite{das_anonymity_2018}.
Most of the work from the second cluster offer probabilistic guarantees via differential privacy~\cite{dwork_calibrating_2006}, which makes meaningful longitudal privacy challenging as privacy degrades with each iteration.

The third clusters include designs focusing on low latency, using mixnets and onion routing.
None of these offer any provable privacy guarantees, and e.g. Tor~\cite{dingledine_tor_2004} has been shown to be vulnerable to many deanonymization attacks~\cite{karunanayake_-anonymisation_2021}.

The last cluster consists mainly of cover protocols and steganography techniques.
Most of these work have strong trust assumptions, and weaker adversaries than the work that does not offer strong censorship resilience.
Unfortunately, none of them provide any provable privacy guarantees on the traffic shape of the protocol acting as the tunnel.
In contrast, the work in this paper maintains the traffic shape of the tunneling ('regular') protocol by design, which we prove in \Cref{thm:denim:privacy}.

\ourprotocol is the only work that combines both strong censorship resilience with provable privacy guarantees.
Moreover, this work is the only that models the information flows on the application layer when constructing the proof. 
With our instantiation \ourprotocol, the simplicity of the design comes at the cost of a centralized server.
The closest work is by Howes~IV~et~al.~\cite{iv_security_2022}, which proposes a framework to formalize and analyze tunneled traffic on the transport layer, to protect against global passive adversaries. 

\ourparagraph{Relation to work on the Signal protocol}
Initial work related to Signal included formalizing the protocol in terms of cryptographic primitives and proving its security under minimal trust assumptions \cite{cohn2020formal,bellare2017ratcheted,jaeger2018optimal,alwen2019double}. 
Over time the attention has shifted towards other aspects as well, including implicit vs robust authentication \cite{blazy2019said,blazy2022marshal}; and offline cryptographic deniability \cite{unger2018improved,vatandas2020cryptographic,brendel2022post}. 
The latter line of research is closest to this work. The state-of-the-art on this matter is that, from the cryptographic perspective, the Signal protocol only provides offline deniability \cite{vatandas2020cryptographic} (transcripts provide no
evidence even if long-term key material is compromised); but no online deniability \cite{unger2018improved} (outsiders can obtain evidence of communication). We bypass this impossibility result by assuming receivers of deniable messages to be honest, and making deniable messages unobservable at a network level, thus strengthening the deniability claims achieved by \ourprotocol\ in comparison to standard Signal.

\section{Conclusion}\label{sec:conclusion}
This work introduces \denimonsignal, an instant messaging system that provides different privacy guarantees for messages in the same system: either with or without metadata privacy.
We show that it is possible to design and run a state-of-the-art, stateful protocol that guarantees provable metadata privacy.
Specifically, we design a variant of the Signal protocol, \ourprotocollong, and show that subtle changes -- all caught by a formal model -- to the original protocol are necessary for privacy.
Through our proof-of-concept implementation \denimonsignal we showcase how to strike performance trade-offs for real-world applications -- the overhead of deniable traffic in \ourprotocol\ is parameterizable through a variable \padding, and we empirically show the behavior under different traffic loads.


 \bibliographystyle{IEEEtran}
 \bibliography{references.bib}

\iffullversion
\onecolumn
\appendix

\fi
\fi

\end{document}